\documentclass[sigplan,nonacm,screen]{acmart}

\usepackage{algpseudocode}
\usepackage[ruled,vlined,linesnumbered,noline,noend]{algorithm2e}
\usepackage{booktabs,tabularx,array,pifont,threeparttable}

\SetKwInput{KwIn}{Input}
\SetKwInput{KwOut}{Output}
\SetKwProg{Fn}{Function}{}{}

\usepackage{enumitem}
\usepackage{amsmath}
\usepackage{xcolor}
\usepackage{listings}
\usepackage{float}

\lstdefinelanguage{mypython}{
  language=Python,
  morekeywords={Exp,parfor,Load,Store,MMA,Rowsum,Div,MMA_transB},
}

\lstdefinestyle{customc}{
  numbers=left,
  numberstyle=\footnotesize,
  numbersep = 2pt,
  xleftmargin=5pt, %
  belowcaptionskip=1\baselineskip,
  breaklines=true,
  language=mypython,
  showstringspaces=false,
  abovecaptionskip=0pt,
  belowcaptionskip=0pt,
    aboveskip=0pt,
    belowskip=0pt,
  basicstyle=\linespread{0.2}\footnotesize\ttfamily,
  keywordstyle=\bfseries\color{green!40!black},
  commentstyle=\itshape\color{purple!40!black},
  identifierstyle=\color{blue},
  stringstyle=\color{orange},
}
\floatstyle{plain}
\newfloat{lstfloat}{t}{lop}
\floatname{lstfloat}{Listing}

\usepackage{soul}
\usepackage{xspace}

\newcommand{\eg}{\textit{e.g.,}\xspace}
\newcommand{\ie}{\textit{i.e.,}\xspace}
\newcommand{\mypara}[1]{%
  \vspace{0.05in}
  \noindent\textbf{#1}
}

\newcommand{\splitk}{Split-K\xspace}
\newcommand{\itgraph}{itGraph\xspace}
\newcommand{\itgraphs}{itGraphs\xspace}
\newcommand{\algebraic}{algebraic\xspace} %
\newcommand{\tmap}{\texttt{tmap}\xspace}
\newcommand{\Split}{Split\xspace}

\newcommand{\sysname}{\textsc{Cleave}\xspace}
\newcommand{\tilelang}{TileLang\xspace}
\newcommand{\Welder}{Welder\xspace}
\newcommand{\Mirage}{Mirage\xspace}
\newcommand{\taso}{TASO\xspace}
\newcommand{\Neptune}{Neptune\xspace}
\newcommand{\Trinity}{Trinity\xspace}
\newcommand{\flashattention}{FlashAttention\xspace}
\newcommand{\flashdecoding}{FlashDecoding\xspace}

\newcommand{\avgOpCompSpeedupOvrMirage}{5.9$\times$\xspace}
\newcommand{\maxOpRunSpeedup}{2.8$\times$\xspace}
\newcommand{\maxModelSpeedup}{2.5$\times$\xspace}
\newcommand{\avgOpRunSpeedup}{1.6$\times$\xspace}
\newcommand{\avgModelSpeedup}{1.7$\times$\xspace}
\newcommand{\splikimprovement}{2.1\xspace}
\newcommand{\maxOpRunSpeedupDynamicFAII}{1.4$\times$\xspace}
\newcommand{\maxOpRunSpeedupDynamicFAIII}{1.7$\times$\xspace}

\newcommand{\codelight}[1]{\hl{#1}}

\usepackage[font={small},labelfont={small,bf},textfont={small},skip=2pt]{caption}
\usepackage{subcaption}

\newcommand{\xmark}{\textcolor{red}{\ding{55}}}

\begin{document}

\title{%
\textsc{Cleave}: Scaling Tensor Program Optimization via Decoupled Algebraic Search and Operator Scheduling}

\author{David Pissarra}
\authornote{Equal contribution.}
\affiliation{%
  \institution{New York University}
  \city{New York}
  \country{USA}}
\email{david.pissarra@nyu.edu}

\author{Jinkun Lin}
\authornotemark[1]
\affiliation{%
  \institution{Cornell University}
  \city{Ithaca}  %
  \country{USA}}
\email{jinkun.lin@nyu.edu}

\author{Haitian Jiang}
\affiliation{%
  \institution{New York University}
  \city{New York}
  \country{USA}}
\email{haitian.jiang@nyu.edu}

\author{Aurojit Panda}
\affiliation{%
  \institution{New York University}
  \city{New York}
  \country{USA}}
\email{apanda@cs.nyu.edu}

\author{Jinyang Li}
\affiliation{%
  \institution{New York University}
  \city{New York}
  \country{USA}}
\email{jinyang@cs.nyu.edu}

\renewcommand{\shortauthors}{Pissarra et al.}

\begin{abstract}
Optimized kernels such as FlashAttention and FlashDecoding are crucial for  accelerating today's large models. Most of them are handwritten by experts  because existing ML compilers cannot match their efficiency. Producing such kernels requires fusing computations with multiple reductions, which requires both algebraic transformation of the computation graph and operator scheduling of the transformed graph. Unfortunately, searching the two jointly yields a space too large to navigate. We propose \sysname, an ML compiler built on \emph{symbolic decoupling}: \sysname discovers transformations by performing superoptimization on a graph with symbolic shapes, and then schedules each resulting graph on concrete shapes. Representing shapes as symbols makes equivalence checking cheap and lets a new Split operator, with a symbolic split count, parallelize along a reduction dimension. \sysname's scheduler fuses graphs with multiple reductions through iterative tiling and horizontal fusion. Evaluation on common LLM subgraphs shows that \sysname generates kernels up to \maxOpRunSpeedup faster than the best baseline (\avgOpRunSpeedup on average) and reduces compilation time by \avgOpCompSpeedupOvrMirage on average compared to \Mirage. For dynamic workloads captured from production serving traces, \sysname
compiles each operator once and achieves geometric mean
speedups of \maxOpRunSpeedupDynamicFAII and \maxOpRunSpeedupDynamicFAIII over FlashInfer's handwritten FA2 and FA3
backends.
\sysname's code is available at: \url{https://github.com/nyu-systems/cleave}.

\end{abstract}

\maketitle

\section{Introduction}
\label{sec:introduction}

\begin{figure}[t]
    \centering
    \includegraphics[width=0.49\textwidth]{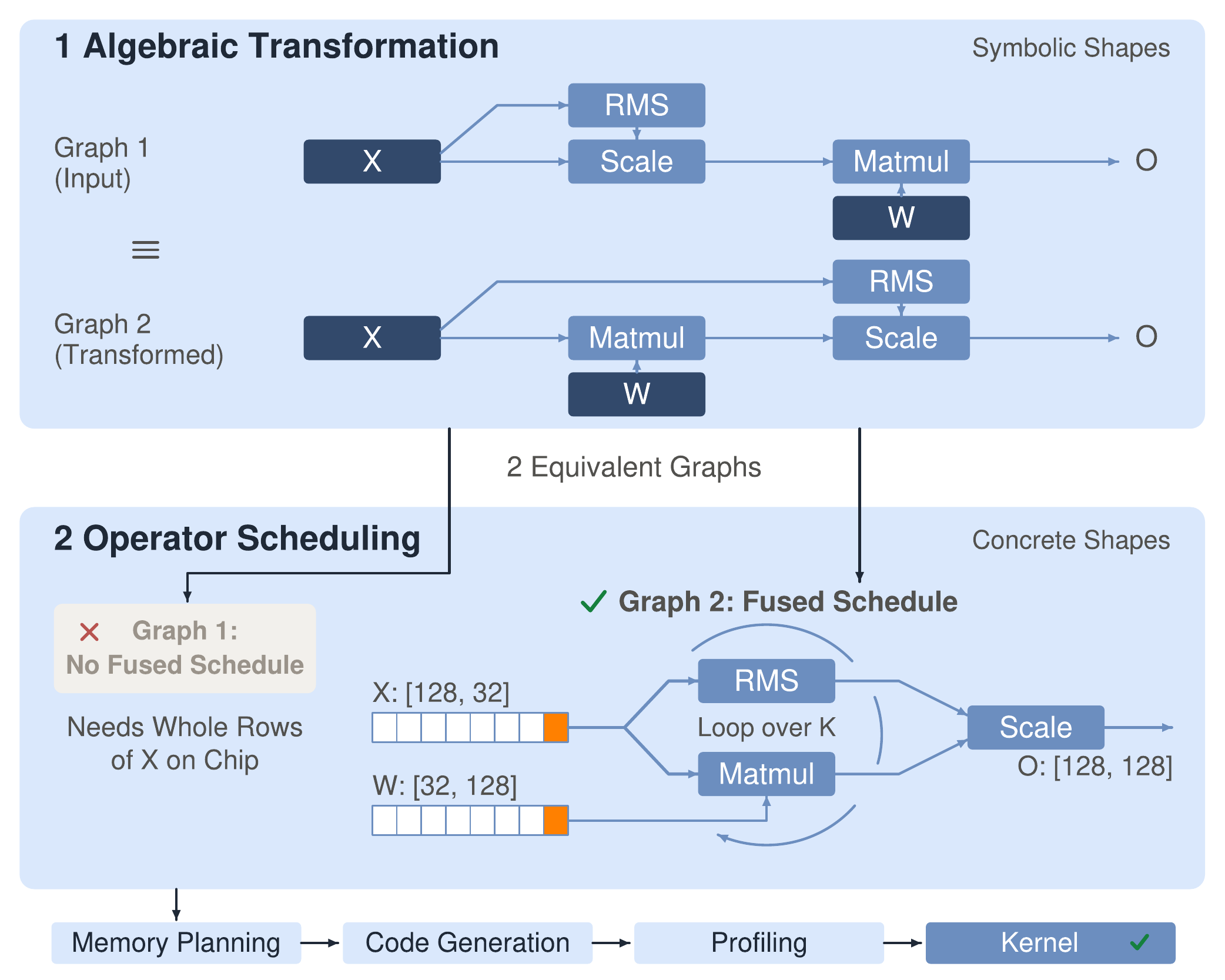}
    \caption{Symbolic decoupling in \sysname, illustrated by the fusion of RMSNorm followed by Matmul. The first phase performs algebraic transformation that finds an equivalent graph which orders the row-wise scaling after Matmul. The second phase performs operator scheduling which makes each slice of $X$ feed both reductions in a single loop over $K$.}
    \label{fig:decoupling}
\end{figure}

Efficient execution of deep learning models requires high-performance GPU kernels, which must minimize data movement and exploit hardware parallelism. Kernel fusion is an important optimization technique for reducing data movement: it combines multiple operations into one kernel so that intermediate results can be kept in shared memory or registers, thereby avoiding global memory traffic. Building efficient fused kernels, as in FlashAttention~\cite{flashattention,flashattention3} and FlashDecoding~\cite{flashdecode}, has required substantial effort from human experts. Automating this work would reduce the effort needed to optimize new models. However, despite extensive work on ML compilers~\cite{mcfuser,welder,ladder,chimera,ansor,astitch,rammer,korch,taso,einnet,pet,egg-mlsys,flashtensor,dnnfusion,mirage}, automatically generating fused kernels that match the efficiency of expert implementations remains difficult.

Kernel fusion is difficult for modern deep learning models
because of their extensive use of various reduction-based operations, including matrix multiplication, convolution, normalization and softmax. Fusing these computations often requires combining multiple reductions in one kernel. However, dependencies between reductions can require more
intermediate storage than shared memory and registers provide. Furthermore, the straightforward way of partitioning the output across threads may expose too little parallelism to use the GPU efficiently.

Resolving these challenges requires more than determining how to tile, order and map the computation to GPU threads, a process we refer to as \emph{operator scheduling}. It also requires \emph{algebraic transformation} to restructure the computation graph to enable tile-by-tile execution and expose additional parallelism. For example, when RMSNorm is followed by a Matmul (Fig.~\ref{fig:decoupling}), moving the row-wise scaling of RMSNorm after the Matmul allows both reductions to consume the same input tile together, avoiding materialization of the normalized input. The same transformation, applied to the softmax normalization, underlies \flashattention (\S\ref{s:bg:case}). Scheduling then determines how to tile and execute the transformed graph. Likewise, when a reduction produces too few output tiles to occupy the GPU, algebraic transformation can decompose it into independent partial computations whose results are subsequently combined, a technique commonly referred to as \splitk. Scheduling then determines the partition sizes and maps the partial computations across threadblocks, as in \flashdecoding.

Existing compilers address these requirements in different ways. Scheduling-based compilers~\cite{mcfuser,welder,ladder,chimera,ansor,astitch,rammer} optimize loops, tile sizes, and memory layouts, but do not discover the algebraic transformations needed above. Algebraic optimizers~\cite{korch,taso,einnet,pet,egg-mlsys,flashtensor,dnnfusion} search for equivalent computation graphs, but graph transformations alone do not determine how to tile and execute them. Mirage~\cite{mirage} uses superoptimization to jointly search algebraic transformations and operator schedules, resulting in a search space that is often too large to optimize. As our evaluation (\S\ref{subsec:operator-val}) shows, Mirage fails to optimize some commonly used kernels and runs out of memory on some searches. Beyond compilers, a separate line of work uses LLM agents to write kernels directly~\cite{kernelbench,accelopt,kernelevolve,ksearch,cake}. These agents match expert kernels on some workloads, but each new operator requires a costly new search.

We observe that, when combining algebraic transformation with scheduling for fusion, the underlying optimizations can be separated along a natural line. Algebraic transformation rewrites the computation graph into an equivalent one, changing the operators and the dependencies among them while preserving the final result. Scheduling leaves the graph unchanged and decides how it is executed: how reductions are tiled into loops, which loops are fused and how tiles are mapped to threadblocks. The validity of a transformation is a property of the graph alone and does not depend on the schedule nor the tensor shapes. Thus, it can be established once for a graph with symbolic shapes. Whether a transformed graph is profitable depends on the schedule and concrete shapes, which can be decided by a scheduler. We call this approach \emph{symbolic decoupling}: the algebraic phase searches over graphs with symbolic dimensions, and the scheduling phase binds the symbols to concrete shapes and schedules each resulting graph. In effect, the algebraic search never enumerates loop structures and tile sizes, and the scheduler never reasons about algebraic equivalence.%

We have implemented symbolic decoupling in \sysname.  At a high level, \sysname first uses superoptimization to produce all algebraically equivalent graphs of the input and then uses a dedicated scheduler
to perform fused tile-by-tile execution for each graph. \sysname chooses the most promising candidates according to an analytical cost model and compiles and profiles them for the final kernel. In order for this approach to work well, \sysname addresses two technical challenges:
\begin{enumerate}[nosep, noitemsep,leftmargin=*]
    \item \emph{Enhancing and scaling algebraic search.} (\S\ref{sec:algebraic-transformation}) Given the principle of symbolic decoupling, how to enable the algebraic transformation required to support \splitk? To do so, we introduce a new Split operator in the symbolic graph that partitions a reduction dimension into $s$ parts, where $s$ is one more symbol that divides a given tensor dimension. Superoptimization then decides where to insert Split and how to combine the partial results, while the concrete value of $s$ is bound later by the scheduler. The symbolic graph representation also makes the search scale. Superoptimization checks the equivalence of each candidate graph by probabilistic testing~\cite{mirage}, whose cost grows with the tensor sizes. For symbolic graphs, we can perform probabilistic testing using any small instantiation that
    satisfies the graph's shape constraints.
    Note that at the end of superoptimization, we still perform the probabilistic testing on the original shapes from the user graph to not lose the correctness guarantee.

    \item \emph{Scheduling graphs with multiple reductions.} (\S\ref{sec:scheduling}) An algebraically transformed graph still needs a schedule that keeps intermediate storage small and reuses data across reductions. Treating reductions as kernel boundaries, or materializing their producers' full intermediate tensors, would lose the fusion opportunities exposed by algebraic search. Our scheduling algorithm combines two strategies: (1) \emph{iterative tiling} to break reduction dimensions into smaller tiles, each of which can be computed without materializing all intermediate results in shared memory; and (2) \emph{horizontal fusion} to compute reductions that share upstream producers in the same loop, reusing the tiles those producers generate.
\end{enumerate}

\sysname also supports dynamic workloads (\S\ref{subsec:dynamic}), where batch size and sequence length change from one invocation to the next. %
In this case, scheduling is performed on a representative concrete shape, and code generation keeps the dynamic extents symbolic with runtime loop bounds, so one compiled kernel serves the varying request shapes. %

We have evaluated \sysname on commonly used subgraphs, including GQA, MLA, and SwiGLU. Our evaluation (\S\ref{sec:evaluation}) shows that kernels optimized by \sysname are up to \maxOpRunSpeedup faster than the best baseline on each subgraph, with an average speedup of \avgOpRunSpeedup. Compared to \Mirage, \sysname reduces compilation time by \avgOpCompSpeedupOvrMirage on average. On dynamic workloads, a single compilation per operator serves all 303 captured shapes across nine operators, achieving geometric mean speedups of \maxOpRunSpeedupDynamicFAII and \maxOpRunSpeedupDynamicFAIII over FlashInfer's FA2 and FA3 backends, respectively. Finally, when \sysname-produced kernels are used in transformer layers from GPT-2, Llama-3 8B, and Qwen3-32B, layer execution improves by up to \maxModelSpeedup (\avgModelSpeedup on average) over \texttt{torch.compile}.

\section{Motivation and Our Approach}

We start by presenting some background on kernel optimization (\S\ref{sec:terminology}); then show, through \flashattention and \splitk, that fusing reduction-based computations requires both algebraic transformation and operator scheduling (\S\ref{s:bg:case}); and finally present the two techniques \sysname uses to meet these requirements, symbolic decoupling and iterative tiling (\S\ref{s:bg:obs}).

\begin{figure}[t]
    \centering
    \includegraphics[width=0.49\textwidth]{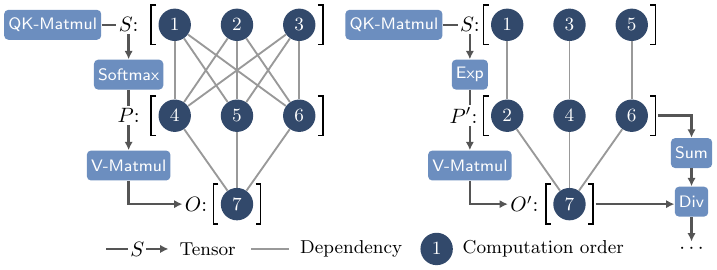}
    \caption{Data-dependency in the attention graph for computing one output element, before (left) and after (right) \flashattention algebraic transformation. Circles represent elements in the tensors. Labels indicate computation order after fusion. Before, $S$ is fully computed and cached in SMEM before computing $P$, incurring high SMEM usage. After, $S$ is streamed with $P'$, reducing SMEM usage.}
    \label{fig:attn-dependency}
\end{figure}

\subsection{Background}
\label{sec:terminology}

\mypara{GPU execution and memory model.} We use NVIDIA's terminology throughout the paper. A kernel is executed by many threads organized into \emph{threadblocks}, each of which runs on one streaming multiprocessor. Threads have access to three tiers of memory with each level larger but slower than the previous one: thread-local registers (RMEM), shared memory (SMEM) accessible by all threads in a threadblock, and global memory (GMEM) accessible by all threadblocks. A kernel commonly partitions its output tensor into tiles and assigns one threadblock to compute each tile. Thus, the number of output tiles determines the degree of parallelism.

\mypara{Tiling and fusion.} To improve kernel performance, one must reduce data movement. \emph{Tiling} breaks a computation into blocks small enough to fit in SMEM or RMEM, so that data read from GMEM can be reused across the block rather than reloaded. \emph{Kernel fusion} merges consecutive kernels into one so that intermediate results stay in SMEM or RMEM instead of being written to GMEM and read back across kernel boundaries.

\mypara{Superoptimization.} Superoptimization~\cite{taso,mirage} finds transformations of a computation graph by enumerating candidate graphs up to a threshold size and checking for those semantically equivalent to the input graph. %
Equivalence between a candidate graph and the input can be established by probabilistic testing, which runs both on random inputs and compares their outputs~\cite{mirage}. The cost of each test grows with the sizes of the tensors involved.

\subsection{Motivating Examples}\label{s:bg:case}

We use two optimizations, \flashattention and \splitk, to show that optimizing reduction-based computations requires both algebraic transformation and operator scheduling. Both serve as running examples in the rest of the paper.

\begin{figure}[t]
    \centering
    \includegraphics[width=0.49\textwidth]{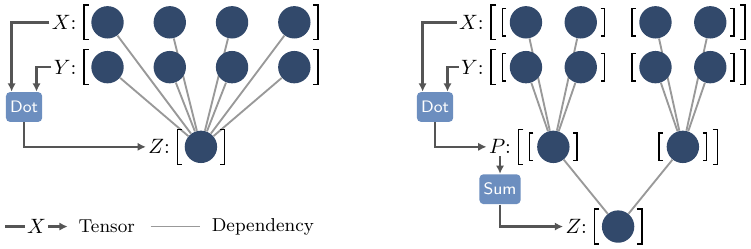}
    \caption{Algebraic transformation in \splitk, which rewrites the vector dot product $X\cdot Y$ into $X[:2]\cdot Y[:2]+X[2:]\cdot Y[2:]$.}
    \label{fig:splitk-dot}
\end{figure}

\mypara{\flashattention.}
At its core, the attention operation of transformer models consists of two matrix multiplications (a QK-Matmul and a V-Matmul) with softmax sandwiched in between.
Optimizing the original attention graph into \flashattention requires two steps: first, the computation graph needs to be transformed into the equivalent graph where the division operator of softmax is after matrix multiplication (V-Matmul). Next, the resulting graph needs to be scheduled in a way so that execution proceeds tile by tile along the reduction dimension of V-Matmul, \ie the sequence dimension. Each iteration computes one tile of $S$ and $P'$ and accumulates its contribution to the output, so only one tile of each intermediate tensor is live at a time.

Importantly, tiling along the reduction dimension before transforming the graph is ineffective. As shown in Fig.~\ref{fig:attn-dependency}, the attention graph contains a problematic reduction operator (softmax), whose input and output tensors are both large. Fusing it according to the original graph requires substantial memory because the entire $S$ must be materialized in SMEM or RMEM before computing $P$, as seen by the data dependency on the left of Fig.~\ref{fig:attn-dependency}.

Algebraic transformation can eliminate the need to materialize intermediate tensors in their entirety. In the case of softmax, moving the division (part of the softmax computation) after V-Matmul breaks the all-to-all data dependency between $S$ and $P$. As shown in Fig.~\ref{fig:attn-dependency} (right), now each element of $P'$ only depends on a single element of $S$, thereby allowing a schedule to keep one tile of each live at a time instead of entire tensors.

\mypara{\splitk.}
Since the number of output tiles determines the degree of parallelism (\S\ref{sec:terminology}), a reduction with a small output yields too few threadblocks to fully occupy the GPU. This is common in decode attention, where each request contributes only a single row of the output $O$ tensor.
\splitk~\cite{splitk} improves parallelism by partitioning the reduction dimension. The example of Fig.~\ref{fig:splitk-dot} applies \splitk to the dot product $Z=\mathrm{Dot}(X,Y)$ of length K (left). The result is $Z=\sum_{j=0}^{s-1}\mathrm{Dot}\bigl(X[jb:(j+1)b],\,Y[jb:(j+1)b]\bigr)$ with $b=K/s$ (Fig.~\ref{fig:splitk-dot}, right). The $s$ partial dot products can now run in parallel on separate threadblocks with an extra \emph{combine} kernel summing their results. Here, algebraic transformation and scheduling appear to be coupled in the opposite order from \flashattention: the rewrite is stated in terms of the partition size $b$. As choosing partitions is a scheduling decision, it appears that the schedule must be fixed before the transformation can be applied.

\subsection{Our Approach}\label{s:bg:obs}
Existing work considers either operator scheduling only~\cite{mcfuser,welder,ladder,chimera,ansor,astitch,rammer} or algebraic transformation only~\cite{korch,taso,einnet,pet,egg-mlsys,flashtensor}.
Mirage~\cite{mirage} performs joint optimization of both but suffers from a blow-up of the search space.
Our system \sysname performs both algebraic transformation and operator scheduling using two techniques. \emph{Symbolic decoupling} performs algebraic transformations independently of scheduling. \emph{Iterative tiling} then schedules each transformed graph so that its reductions are fused without materializing intermediate tensors.

\mypara{Symbolic decoupling.} Symbolic decoupling rests on a distinction between a computation graph and its execution. The algebraic phase transforms the graph into an equivalent one, changing the operators, their ordering, and the dependencies among them while preserving the result. The scheduling phase does not change the graph but decides how it executes: how reductions are tiled into loops, which loops are fused, and how tiles map to threadblocks. The validity of a transformation is a property of the graph alone and should be independent of the schedule and of the tensor shapes. Thus, it can be established on a symbolic graph before any scheduling decision is made. Every quantity the transformation does not depend on, e.g. every tensor's dimension, is represented by a symbol. %
We call this \emph{symbolic decoupling}. The two examples of \S\ref{s:bg:case} readily admit this approach as follows.

The \flashattention transformation, which moves the division after the V-Matmul, changes the graph and nothing else. It is valid for every schedule and every sequence length and model dimension. The algebraic phase can discover this transformation. The scheduling phase then tiles the computation according to the transformed graph without further algebraic reasoning.
The \splitk rewrite also changes only the graph, and it is valid for every partition size. The apparent coupling comes from stating it with a concrete $b$. Under symbolic decoupling, the number of parts $s$ is simply one more symbolic dimension: the algebraic phase introduces it through a Split operator that partitions a dimension into $s$ parts, and decides where to split and how to combine the partial results. The scheduling phase resolves $s$ along with the other symbols (\S\ref{sec:algebraic-transformation}).

\mypara{Iterative tiling.} Algebraic transformation creates the opportunity to fuse a graph with multiple reductions, but we still need to design a scheduler to realize it. To fuse such a graph without materializing intermediate tensors, the scheduler must tile each reduction along its reduction dimension and accumulate the result tile by tile so only one tile of each intermediate is live at a time. Furthermore, when several reductions consume the same producer, one must compute them in the same loop, so that each tile of the producer is computed once and reused. We refer to the former as \emph{iterative tiling} and the latter as \emph{horizontal fusion}. The transformed attention graph is one instance: the row sum and the V-Matmul both reduce over the sequence dimension and both consume $\exp(QK^T)$. Existing scheduling-based compilers~\cite{welder,ladder,astitch} do not produce such schedules. They tile only the output dimensions of a kernel, so even when given the transformed graph, they materialize the full intermediate tensors in SMEM, which limits fusion to short sequence lengths for attention. %

\section{System Design Overview}

\sysname optimizes the kernels corresponding to a computation graph. This graph is usually a subgraph of the overall computation graph for end-to-end ML model execution.
\sysname works in two phases: \emph{algebraic transformation} (\S\ref{sec:algebraic-transformation}) and \emph{operator scheduling} (\S\ref{sec:scheduling}). A final pass (\S\ref{sec:codegen-phase}) performs memory placement, generates code for the kernels, and profiles their performance.

First, \sysname symbolizes the graph by replacing original tensor shape dimensions with symbols, uses superoptimization to find all mathematically equivalent ones, validates with probabilistic testing, and instantiates them with original shapes.
Second, operator scheduling transforms the validated graph's loop structure using \emph{iterative tiling} and \emph{horizontal fusion}, and then uses an analytical cost model to find the top-k tile sizes.
Finally, for the top-k candidate graphs, \sysname decides the tensor placement (in SMEM or RMEM) and layout, transpiles into \tilelang to generate the kernels, which are then profiled to pick the best among the candidates.

\section{Algebraic Transformation Phase}
\label{sec:algebraic-transformation}

The algebraic transformation phase takes a computation graph with concrete tensor shapes as input. It creates a corresponding symbolic graph, then performs superoptimization with probabilistic testing, and finally recovers the original shape. The phase returns a set of algebraically-equivalent graphs with concrete shapes.

\mypara{Creating symbolic graphs with support for Split-K.}
We turn the concretely-shaped input computation graph into a symbolic one by representing each tensor dimension with a symbol.
Doing so can greatly reduce the cost of superoptimization because
checking for graph equivalence is commonly done using random testing~\cite{taso,mirage} whose cost is proportional to the tensor size.
With symbolic shapes, we can  check equivalence by running random tests for small tensor sizes.
To create such a graph, we first keep only dimensions with size $1$ to preserve broadcasting structures, and replace all other dimensions with symbolic variables. We then traverse all operators in the graph to capture all shape constraints among the variables, which will be used in the numerical testing when instantiating a small-shape graph.
The lower-level implementation is detailed in \S\ref{sec:implementation}.

As discussed in \S\ref{s:bg:obs}, we introduce the Split operator in the symbolic graph to enable \splitk optimization.
The Split operator splits a given dimension into $s$ parts (where $s$ is a symbolic variable).
For example, $\operatorname{Split}(A: [m,k],\mathrm{dim=1})$ yields a tensor of shape $[s,m,k/s]$.
Using a symbolic variable $s$ instead of a concrete value allows the algebraic transformation phase to focus its search on identifying where tensors should be split without searching for split size.

\begin{figure}
\centering
\begin{subfigure}[b]{1\linewidth}
    \centering
    \includegraphics[width=1\linewidth]{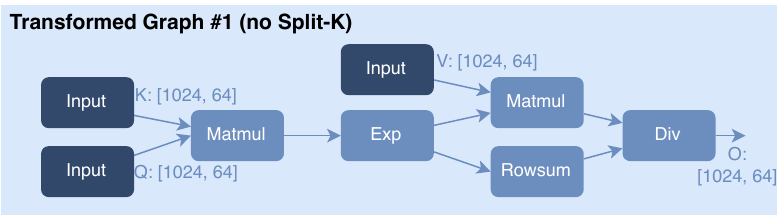}
    \caption{An equivalent attention graph without \Split operators}
    \label{fig:algebra-output-1}
\end{subfigure}
\begin{subfigure}[b]{1\linewidth}
    \centering
    \includegraphics[width=1\linewidth]{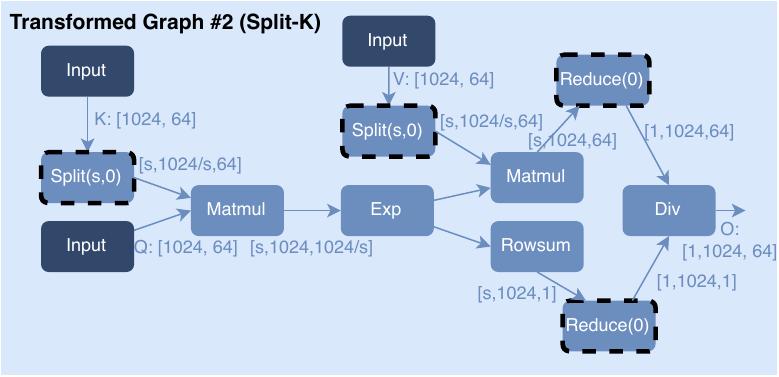}
    \caption{An equivalent attention graph with \Split operators. \Split adds a leading dimension $s$ which is treated as a batch dimension by subsequent operators such as Matmul (which uses broadcasting semantics for inputs with a batch dimension). The symbolic $s$ will be instantiated into valid concrete integers that divide $1024$.
    \label{fig:algebra-output-2}}
\end{subfigure}
\caption{Two equivalent attention graphs discovered via superoptimization, with and without \Split nodes. }
\label{fig:algebra-output}
\end{figure}

When the Split operator is included in the graph, it naturally yields a
two-kernel
decomposition consisting of a \emph{split} kernel and a subsequent \emph{combine} kernel by partitioning at the reduction nodes that reduce the split dimension (\eg the two ``Reduce(0)'' nodes in Fig.~\ref{fig:algebra-output-2}). In attention operators, splitting the key-value reduction dimension produces a \emph{split} kernel that emits two partial tensors, a \emph{per-split} row-sum used for normalization, and a \emph{per-split} value-product tensor of shape \([s,S_Q,D]\). The \emph{combine} kernel then reduces both tensors across \(s\) and performs the final division, yielding the normalized attention output.
Figure~\ref{fig:algebra-output-2} shows this example.

\begin{figure}
    \centering
    \includegraphics[width=1\linewidth]{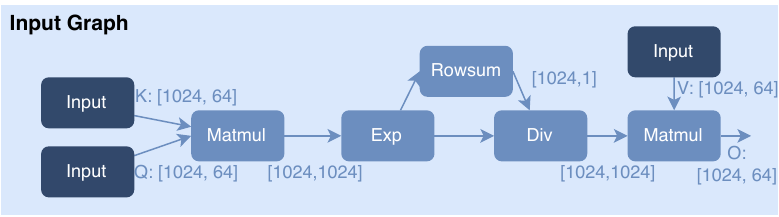}
    \caption{Input to the superoptimizer describing attention computation. The first Matmul uses a transposed layout for the $K$ matrix.}
    \label{fig:algebra-user-graph}
\end{figure}

\mypara{Superoptimization with reducibility-based pruning.}
To speed up superoptimization's enumeration-based search, \sysname prunes the search space by imposing constraints on what dimensions can be split and reduced for \Split and any reduction operators. We first identify for every input tensor of the graph which dimensions are \emph{eventually-reduced}, defined as whether there exists an element in the graph's output tensor such that all elements that contribute to its computation cover all indices in this dimension.
The main idea is that, since any algebraically equivalent graph should preserve the same set of dimensions that are eventually-reduced, \sysname prunes the search path whenever the insertion of \Split or any reduction operators breaks this property. In other words, we never allow splitting or reducing non-eventually-reduced dimensions, \eg batch dimensions. Note that we prune for \Split insertions as in typical \splitk optimization, the split dimensions will be eventually-reduced in the second kernel, making it similar to a reduction operator.
Currently, we require the user to annotate the input tensors for which dimensions are eventually-reduced, but it can be automated with data-dependency analysis~\cite{flashtensor,welder}.

The superoptimization algorithm takes the algebraic graph as input, and enumerates all possible graphs up to a predefined size. The process starts from a disconnected graph with only the input tensors from the algebraic graph and gradually inserts new nodes. To test equivalence, \sysname instantiates small concrete shapes that conform to the operator constraints and performs numerical testing using \Mirage's probabilistic testing approach (details in \S\ref{sec:implementation}).

\mypara{Recovering the original shapes.} For each equivalent graph discovered by superoptimization, we can recover concrete tensor shapes by propagating concrete dimension values starting from the original concrete input tensor shapes. If a symbolic \Split operator is present in the concretized graph, we enumerate all valid possible split values, which are the factors of the corresponding concrete split dimension value, and test for their validity.
Doing so results in several concrete graphs, one for each valid split value. Finally, each \splitk graph is partitioned into a split and combine subgraph, each of which is independently scheduled and generates its own kernel. At runtime, both kernels need to run.

Fig.~\ref{fig:algebra-output} shows two example equivalent graphs discovered via superoptimization for the attention graph in Fig.~\ref{fig:algebra-user-graph}.

\section{Operator Scheduling Phase}
\label{sec:scheduling}

\subsection{Motivating Example and Overview}
\label{subsec:sched-motivating-ex}

\begin{lstfloat}
\begin{lstlisting}[language=mypython, caption={Tile program pseudocode for \flashattention schedule. Each threadblock computes one row slice of $O$ as highlighted.\label{lst:fa}},tabsize=2]
# seqlen (N), head_dim (d), output & reduction tile size (b_O & b_R)
parfor i in range(0, N, b_O): # parallelized across threadblocks @\label{lst:fa:parfor}@
  t_Q = Load(Q[i:i+b_O, :]) # [b_O, d]@\label{lst:fa:load-q}@
  for j in range(0, N, b_R): # iterative tiling@\label{lst:fa:seqfor}@
    t_K = Load(K[j:j+b_R, :]) # [b_R, d]
    t_V = Load(V[j:j+b_R, :]) # [b_R, d]
    t_mma_0 = MMA_transB(t_Q, t_K) # [b_O, b_R]
    t_exp = Exp(t_mma_0) # [b_O, b_R]
    t_rowsum += Rowsum(t_exp) # [b_O, 1]@\label{lst:fa:reuse1}@
    t_mma_1 += MMA(t_exp, t_V) # [b_O, d]@\label{lst:fa:reuse2}@
  t_O = Div(t_mma_1, t_rowsum) # computes row tile O[i:i+b_O:, :]
  Store(t_O, O[@\codelight{i:i+b\_O}, :])@@\label{lst:fa:store}@

\end{lstlisting}

\end{lstfloat}
Ideally, given Figure~\ref{fig:algebra-output-1}, an equivalent transformed graph for attention (returned by superoptimization),
\sysname should produce the code shown in Listing~\ref{lst:fa}. This code uses the \flashattention optimization, and computes a single-head non-causal self-attention (for prefill) on one sequence, with $N,d$ being the sequence length and head dimension, respectively. The program takes $Q, K, V$ as input tensors, and produces output tensor $O$, which all have shape $[N,d]$.

Generating such a program requires applying two scheduling steps.
The first scheduling step %
partitions computation \emph{across threadblocks}.
The program (lines~\ref{lst:fa:parfor}---\ref{lst:fa:store})
partitions the output tensor $O$ into tiles along the row dimension $N$ with tile size $b_O$, and distributes them across threadblocks.
Each threadblock runs the graph to compute one output tile $O[i:i+b_O, :]$.
The kernel uses \texttt{parfor} (L\ref{lst:fa:parfor}) to launch $N/b_O$ threadblocks.
Each threadblock identifies the input slice (\eg $Q[i:i+b_O,:]$) needed for computing their corresponding output tile according to data-dependency, then loads them into SMEM/RMEM (\eg $t_Q$ on L\ref{lst:fa:load-q}) to avoid GMEM round-trips before executing the necessary computation, and finally stores the output tile back to GMEM (L\ref{lst:fa:store}).
\begin{lstfloat}
\begin{lstlisting}[language=mypython,caption={Comparison between schedules with and without iterative tiling. \texttt{g\_mma\_0} and \texttt{g\_exp} are the corresponding full tensors in the computation graph of Fig.~\ref{fig:algebra-output-1}. 
Notice the reduction in the intermediate buffer sizes, as highlighted in the code.
\label{lst:fa-rtile}},tabsize=2]
# without iterative tiling
t_mma_0 = ...      # g_mma_0[i:i+block_O, @\codelight{0:N}@]
t_exp = Exp(t_mma_0) # g_exp[i:i+block_O, @\codelight{0:N}@]@\label{lst:fa-rtile:exp-wo}@
t_rowsum = Rowsum(t_exp)

# with iterative tiling
for j in range(0, N, b_R):
  t_mma_0 = ...     # g_mma_0[...,@\codelight{j:j+b\_R}@]
  t_exp = Exp(t_mma_0) #g_exp[...,@\codelight{j:j+b\_R}@]
  t_rowsum += Rowsum(t_exp) # [b_O, 1]
\end{lstlisting}
\end{lstfloat}

The second scheduling step schedules computation within a threadblock.
This step has two goals: (a) perform \emph{iterative tiling} that tiles reduction nodes to reduce SMEM/RMEM usage; and (b) perform \emph{horizontal fusion}~\cite{bell2024horizontal,ansel2024pytorch} to reuse data.

We illustrate the effect of tiling reduction nodes in Listing~\ref{lst:fa-rtile}.
We denote by $g_{mma,0}$ and $g_{exp}$ the full tensor of the first Matmul and Exp output in the computation graph (Fig.~\ref{fig:algebra-output-1}).
Without iterative tiling, Rowsum would need the entire row of Exp's output $g_{exp}$ to be stored in the SMEM buffer $t_{exp}$ at once, as shown in L\ref{lst:fa-rtile:exp-wo}.
This would require allocating $[b_O, N]$ memory, and lead to suboptimal performance because the allocation size is limited by available SMEM/RMEM.
With iterative tiling, the dimension $N$ is tiled with tile size $b_R$, and iterated with a loop of extent $N/b_R$. Each iteration only sums the column slice $[j:j+b_R]$.
As a result, Rowsum allows staging only these columns of $g_{exp}$ at each iteration, and reuses the same SMEM buffer across iterations. This reduces the buffer size of $t_{exp}$ from $[b_O, N]$ to $[b_O, b_R]$ (L\ref{lst:fa-rtile:exp-wo}).
Similarly, by tracking the data-dependency backwards from Exp to Matmul, we deduce the slice of Matmul's output $g_{mma,0}$ that is needed to compute $t_{exp}$, and stage only that slice in $t_{mma,0}$. In this trivial case where Exp is an elementwise operator, we stage the same column slice $[j:j+b_R]$ of $g_{mma,0}$ in $t_{mma,0}$, reducing the required allocation to $[b_O, b_R]$.
In the program, this is equivalent to moving all of Rowsum's upstream nodes (\ie Matmul and Exp) under $j$'s reduction loop.

Second, the kernel performs horizontal fusion on lines~\ref{lst:fa:reuse1}-\ref{lst:fa:reuse2} in Listing~\ref{lst:fa}. In detail, as the second Matmul is also a reduction sharing the same input $t_{exp}$, we schedule it to compute side-by-side with Rowsum in the same loop, reusing $t_{exp}$.

In summary, the kernel is scheduled by first performing \emph{inter-threadblock scheduling} that partitions the output tensor across threadblocks, and then \emph{intra-threadblock scheduling} that tiles the reduction into sequential accumulation steps with horizontal fusion. The \textbf{goal} of \sysname is to automatically apply the above scheduling steps to any input graph.

\begin{algorithm}[t]
\caption{Overall Workflow of Scheduling Phase}
\label{alg:itgraph_generation}
{
\small

\KwIn{Concrete transformed graph $G_t$}
\KwOut{Top-$k$ itGraph schedules}

\SetKwFunction{EnumTiles}{EnumerateTileSizes}
\SetKwFunction{Schedule}{Schedule}
\SetKwFunction{Cost}{CostModel}
\SetKwFunction{TopK}{SelectTopK}

\tcp{Step 1: enumerate tile size candidates for output and reduction nodes}
$\mathrm{TileConfigs} \leftarrow \EnumTiles(G_t)$\tcp*{\S\ref{sec:implementation}}

\tcp{Step 2: construct itGraphs, filter invalid ones, and evaluate cost}
$\mathrm{Scored} \leftarrow \emptyset$\;
\For{$T \in \mathrm{TileConfigs}$}{\label{alg:itgraph_generation:tileconfig}
    $sch \leftarrow \Schedule(G_t, T)$\tcp*{\S\ref{subsec:itgraph-conversion}}
    \If{$sch.valid$}{
        $c \leftarrow \Cost(sch)$\tcp*{\S\ref{sec:implementation}}
        \If{$c.smem$ fits hardware resources %
        }{
            $\mathrm{Scored} \leftarrow \mathrm{Scored} \cup \{(sch, c)\}$\;
        }
    }
}

\Return{$\TopK(\mathrm{Scored})$}\;
}
\end{algorithm}

\mypara{Overview of the scheduling algorithm.}
The scheduling phase takes as input a concrete transformed graph $G_t$ (e.g., Fig.~\ref{fig:algebra-output-1}), and produces a set of candidate tile programs represented in the \emph{\itgraph} format.
\sysname's approach to doing so is shown in Alg.~\ref{alg:itgraph_generation}:
it first enumerates a set of tile size candidates for output and reduction nodes (\S\ref{sec:implementation}), schedules each candidate into an \itgraph by applying the \emph{inter- and intra-threadblock scheduling} and drops invalid ones (\S\ref{subsec:itgraph-conversion}), and then uses a cost model (\S\ref{sec:implementation})
to select the top few best \itgraphs.
Lower-level \emph{intra-tile} optimizations are delegated to an existing tile program backend~\cite{tilelang}.  Therefore, in the subsequent codegen phase (\S\ref{sec:codegen-phase}), we lower our tile programs into the backend's DSL, compile and profile them, and select the one with the best performance.

In what follows, we first describe the \itgraph semantics (\S\ref{subsec:itgraph}) and discuss how \sysname schedules the computation given the tile sizes into an \itgraph (\S\ref{subsec:itgraph-conversion}).
\sysname adopts an approach similar to \Welder~\cite{welder} for tile size enumeration and uses a similar cost model. We describe both in Appendix~\ref{sec:implementation}.

\begin{figure}[t]
    \centering
     \includegraphics[width=.97\linewidth]{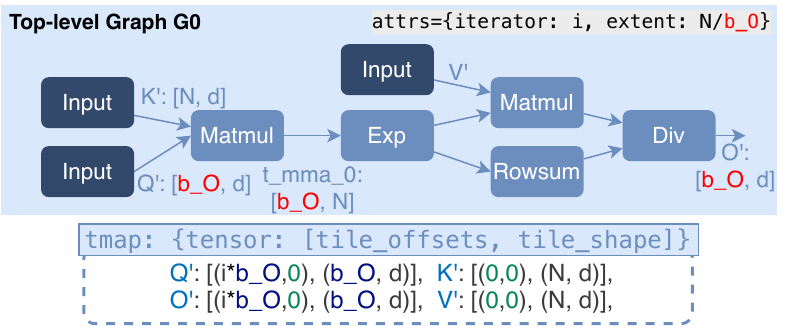}
    \caption{Inter-threadblock scheduling result of Fig.~\ref{fig:algebra-output-1}. The resulting tile program is shown in Listing~\ref{lst:fa-nortile}. It causes high SMEM usage as $t_{mma,0}$ and $t_{exp}$ do not have the second dimension $N$ tiled.
    }
    \label{fig:running-ex-schedule-inter}
\end{figure}

\subsection{\itgraph Semantics}
\label{subsec:itgraph}

Observe that a tile program like Listing~\ref{lst:fa} %
can be regarded as a multi-level loop nest with tile computation and memory IO as its instructions in loop body.
Such a structure naturally maps to a multi-level computation graph, with one level for one loop, and nodes as the instructions operating on tensors. %
We thus enhance the traditional computation graph to support this format, and term it \emph{iterative tile graph} (\emph{\itgraph}).

Formally, an \itgraph describes a sequential or parallel for-loop, and is stored as a computation graph data-structure with attributes of loop variable name and loop extent and a \emph{tensor map} (\tmap). The loop body is defined by the computation graph, indicating that each iteration executes this same graph but using different input and output slices (\ie tiles) in a full tensor. For instance, in Fig.~\ref{fig:running-ex-schedule-inter}, the computation graph of $G_0$ is executed in lines~\ref{lst:fa-nortile:loopbody-begin}---\ref{lst:fa-nortile:loopbody-end} in Listing~\ref{lst:fa-nortile}, but in a parallel loop by all threadblocks, with different slices shown in L\ref{lst:fa-nortile:loopbody-begin} and L\ref{lst:fa-nortile:loopbody-end} each iteration (threadblock).
We use \tmap to encode the slicing formula. The \tmap contains two arrays, both of which have an entry for each dimension: entries in \texttt{tile\_offset} are symbolic expressions that can be used to compute the starting offset of the $i$-th slice of the dimension; and entries in \texttt{tile\_shape} contain the size of each slice.
For example, for $Q[i\times b_O:(i+1)\times b_O, :]$, the \texttt{tile\_offsets} array contains the
symbolic expressions $i\times b_O$ and $0$ for the corresponding dimensions, and the \texttt{tile\_shape} array contains $[b_O, d]$. The symbolic expression $0$ indicates that the corresponding dimension is not tiled.

To allow representing nested loops, we allow \itgraph to nest inner graphs, where nodes in the outer graph can be a \emph{graph-node} that refers to an inner \itgraph that represents an inner loop. In Fig.~\ref{fig:running-ex-schedule-intra}, the graph-node is implicitly shown as the whole $G_1$ graph, which now creates a for-loop on L\ref{lst:fa:seqfor} in Listing~\ref{lst:fa}, containing the execution logic of all nodes in $G_1$.
Every inner graph similarly has \tmap, which slices the tile from the outer graph.
Differently, every inner graph represents a sequential loop, accumulating results to the \emph{same} output tile (\eg $t_{mma,1}$ and $t_{rowsum}$) every iteration. On the other hand, the outermost graph represents a parallel loop where each iteration writes to a different output tile.

\subsection{Inter- and Intra-threadblock Scheduling}
\label{subsec:itgraph-conversion}

Inter- and intra-block scheduling are core to \sysname's approach. In this step, \sysname takes as input a tile size configuration $T$ (Algo.~\ref{alg:itgraph_generation}, L\ref{alg:itgraph_generation:tileconfig}), which contains a tile shape for each of the output and reduction nodes, and produces an \itgraph.
We will use the motivating tile program from \S\ref{subsec:sched-motivating-ex}
 as the running example.
For this example, the tile configuration $T$ contains the tile shape $[b_O, d]$ for output $O$, $b_R$ for both the Matmul and Rowsum nodes, and the full extent $d$ for the first matmul to indicate no tiling. %

\begin{lstlisting}[float, floatplacement=t, language=mypython, caption={After inter-threadblock scheduling, without iterative tiling applied. The corresponding \itgraph is in Fig.~\ref{fig:running-ex-schedule-inter}.\label{lst:fa-nortile}}]
# seqlen (N), head_dim (d), output tile size b_O
parfor i in range(0, N, b_O): # map to threadblocks in parallel
  t_Q = Load(Q[@\codelight{i:i+b\_O}@, :]) # [b_O, d]@\label{lst:fa-nortile:loopbody-begin}@
  t_K = Load(K) # [N, d]
  t_V = Load(V) # [N, d]
  t_mma_0 = MMA_transB(t_Q, t_K) # [b_O, N]
  t_exp = Exp(t_mma_0) # [b_O, N]
  t_rowsum = Rowsum(t_exp) # [b_O, 1]
  t_mma_1 = MMA(t_exp, t_V) # [b_O, d]
  t_O = Div(t_mma_1, t_rowsum) # [b_O, d]
  Store(t_O, O[@\codelight{i:i+b\_O},:])@@\label{lst:fa-nortile:loopbody-end}@
\end{lstlisting}

\begin{figure}[t]
    \centering
    \includegraphics[width=1\linewidth]{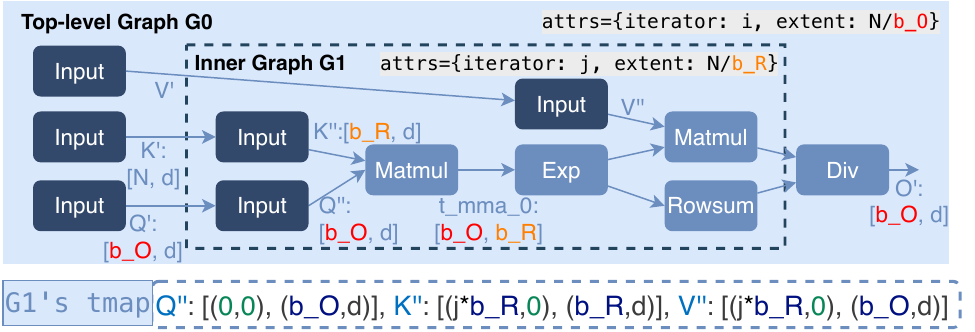}
    \caption{Intra-threadblock scheduling result of Fig.~\ref{fig:running-ex-schedule-inter}, which tiles Rowsum and the second Matmul with $b_R$ size and fuses them together horizontally. $G_1$ is a graph node in $G_0$ and represents an inner-loop. The resulting tile-program is shown in Listing~\ref{lst:fa} (with a minor difference as the tile-program includes changes made by the code generation phase, which hoists $t_Q$ outside the $j$-loop). $G_0$'s \tmap remains unchanged and thus omitted. \tmap of $Q''$ has offsets of $0$ as it uses the same outer $Q'$ across iterations. Notice that the second dimension of $t_{mma,0}$ and $t_{exp}$ is smaller compared to Fig.~\ref{fig:running-ex-schedule-inter}.
    }
    \label{fig:running-ex-schedule-intra}
\end{figure}

Our scheduling algorithm first runs the inter-threadblock pass to partition the output tensor and produces an initial single-level \itgraph (\eg Fig.~\ref{fig:running-ex-schedule-inter}). Next, it repeatedly runs the intra-threadblock passes to create inner-graphs (\eg Fig.~\ref{fig:running-ex-schedule-intra}) on current \itgraph for iterative tiling and horizontal fusion, until all reduction nodes are tiled according to $T$.
We detail this algorithm below. We use ``node'' and ``output tensor'' interchangeably when the node outputs a single tensor.

\mypara{Inter-threadblock scheduling with output tile size.}
\linebreak Given a transformed computation graph $G_t$,
we first build an \itgraph with the same topological structure, calculate its loop extent by dividing the output tensor size by the given tile size, and finally analyze the data-dependency to build the \tmap for input and output tensors.
This analysis starts by constructing the \tmap of the output nodes based on the output tile shape. To calculate \texttt{tile\_offsets} (\ie tile start location at each iteration), we partition the full tensor into disjoint tiles of the given size and iterate them in a linearized order (in row-major).
For example, given output tile shape $[b_O, d]$, with $O$'s full shape as $[N,d]$, its \tmap would have \texttt{tile\_shape} $[b_O,d]$ and \texttt{tile\_offsets} $(i\times b_O , 0)$, where $i$ is the loop variable. %

We use \Welder's data-dependency analysis to propagate the output \tmap and compute \tmap backwards to all other nodes in the graph. Specifically, for each operator we analyze the data-dependency using its tensor expression to deduce the axis range of each of its inputs when its output axis range is constrained to $[0, \text{output\_tile\_size})$. This input range's extent is used as the tile size to construct the \tmap of the input in the same way as in the output \tmap.

During the backward propagation, as multiple paths may exist from the output node back to a node $u$, $u$ can get multiple \tmap{s}. To ensure correctness and for simplicity, whenever any two of $u$'s \tmap do not match exactly, we reject the current tile size candidate and abort its scheduling.%

\mypara{Intra-threadblock scheduling with reduction tile size.}
The schedule produced after the inter-threadblock pass may require large intermediate buffers (\eg $t_{mma,0}$ and $t_{exp}$), resulting in large SMEM usage, because the reduction dimensions (\eg column dimension $N$ of Rowsum) are not tiled and are often large. For example, in attention kernels, they equal the context length which can be thousands or more.
This prohibits graphs with large reduction dimensions from being fused because the intermediate storage required would exceed available SMEM/RMEM,
or prevents the use of larger tile sizes that deliver higher performance.

Our intra-threadblock scheduling applies iterative tiling and horizontal fusion to address this issue, which will break the Rowsum and the second Matmul reduction into multiple smaller steps using their reduction tile sizes $b_R$, and iterate over them in the \emph{same} loop to realize horizontal fusion.

The algorithm first traverses the current \itgraph to identify the ``last-wave'' reduction nodes and separate them into groups for horizontal fusion.
A reduction node is ``last-wave'' of the current graph when it has no other \emph{iterative} reduction nodes, \ie reductions whose extent is large enough to be iterated, in any path from them to the output node, such as Rowsum and the second Matmul in Fig.~\ref{fig:running-ex-schedule-intra} (whereas the first Matmul is not).
We \emph{group two nodes together if they share any of their upstream nodes}, to exploit the reuse of these common nodes in horizontal fusion.
For example, Rowsum and the second Matmul share the Exp node, and thus are grouped together.
Next, we construct one inner \itgraph for each group by wrapping nodes in this group and all their upstream nodes inside, as shown in $G_1$ in Fig.~\ref{fig:running-ex-schedule-intra}.
Finally, we compute \tmap for each inner-graph according to the given reduction tile size. This is done in a similar way as the inter-threadblock pass using tensor expression range-based data-dependency analysis. Note that horizontal fusion requires both fused reduction nodes to iterate over their shared inputs in the same pattern, which we validate by checking \tmap exact match during \tmap backpropagation in the same way as in the inter-threadblock step.

We repeat the same intra-threadblock scheduling on each newly generated inner-graph until all reduction nodes are tiled according to their tile sizes.

\section{Memory Planning and Code Generation}
\label{sec:codegen-phase}

After the scheduling phase, \sysname performs memory placement and layout planning over the resulting \itgraph.
The result is an \itgraph annotated with these decisions. This graph is lowered into \tilelang by emitting buffer allocations, data movement, and tile-level operator calls according to the planned memory and layout decisions.

\subsection{Memory Layout Planning}\label{s:codegen:memory}
\sysname performs unified memory and layout planning over the scheduled \itgraph. In this process, it determines where data should reside, \ie in RMEM or SMEM, based on operator requirements and reuse opportunities; for example, intermediate tensors associated with MatMuls may be placed in SMEM when beneficial to leverage specialized memory instructions. Further, \sysname identifies loop-invariant Input nodes with zero \tmap \texttt{tile\_offset} in inner graphs (\eg the $Q''$ node in Fig.~\ref{fig:running-ex-schedule-intra}) and hoists them to avoid redundant loads from GMEM. During this process, \sysname also maintains tensor core memory layout constraints (Appendix~\ref{apx:tc}).

\subsection{\tilelang Code Generation}
After the memory and layout passes, \sysname lowers the annotated multi-level \itgraph into \tilelang code for kernel generation.\footnote{
We plan to support lowering to other tile-level DSL backends in the future.}
For each level of the \itgraph, \sysname generates two functions:
an iteration function (\texttt{iter}) that implements the loop body, and a wrapper (\texttt{wrap}) that implements the loop by calling the iteration function. Both take an argument per input/output node: the \texttt{wrap} function's inputs are the tensors on which it is operating, while the \texttt{iter} function's inputs are the slices on which it is operating. At each iteration, the wrapper uses the \tmap to determine the arguments it should use when calling \texttt{iter}. Observe that with this separation, the \emph{slicing logic} is implemented in \texttt{wrap} and \emph{execution logic} is implemented in \texttt{iter}.

\subsection{Compiling Dynamic Operators}\label{subsec:dynamic}

Scheduling (\S5) needs concrete tensor dimensions, as it must select tile shapes, thread counts and buffer sizes from them, and its cost model ranks candidates at a fixed shape. \sysname therefore schedules a dynamic operator against one representative concrete shape and then, during code generation, replaces the extents of axes declared dynamic with symbols. Each symbol denotes one dimension of an input tensor, whose size is obtained directly from the tensor at runtime. Every value derived from a dynamic dimension is re-emitted as an expression over its symbol, \ie the launch grid and the size of reduction loops become $\lceil n/b \rceil$ for the symbol $n$ and the corresponding tile size $b$, and existing dynamic memory offsets are computed during runtime on the device.   Because a tile keeps its compile-time size, one that is larger than the input tensor actual size is handled by predicating its loads and stores, and threadblocks left without work are skipped. A single compilation per operator then serves any shape that varies only along its dynamic axes, which are the axes a serving workload changes from one invocation to the next.

\section{Evaluation}
\label{sec:evaluation}

\begin{figure*}[t]
  \centering
  \includegraphics[width=1.0\textwidth]{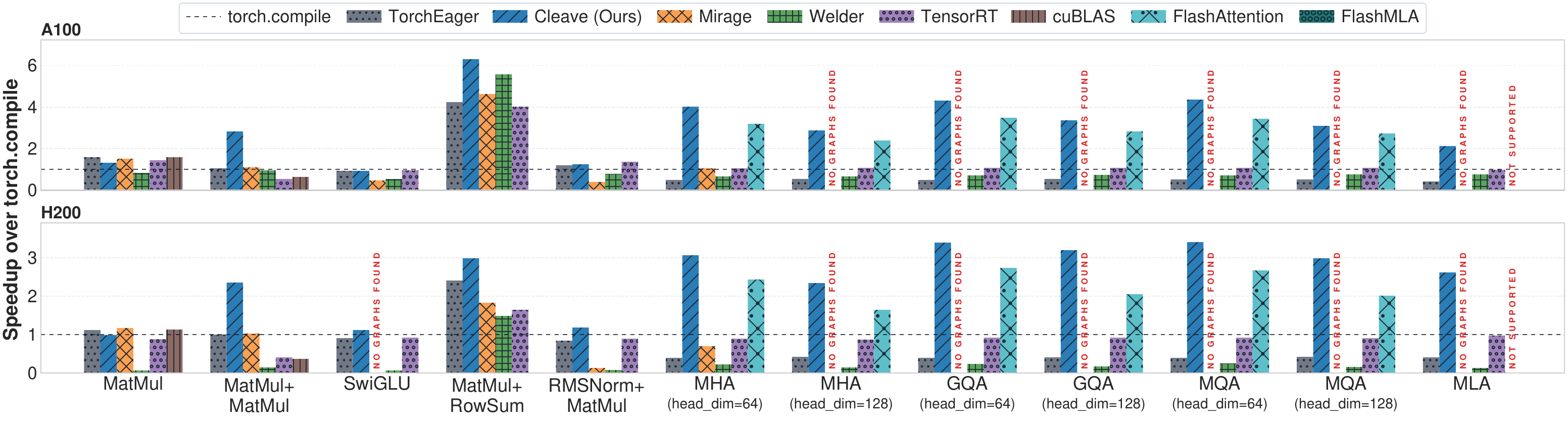}
  \caption{Prefill-based subgraph-level kernel performance speedup against \texttt{torch.compile} with Torch eager mode, Mirage, Welder, TensorRT, cuBLAS, FlashAttention, FlashMLA, and \sysname (ours).}
  \label{fig:prefill_operator_performance}
\end{figure*}

\begin{figure*}[t]
  \centering
  \includegraphics[width=1.0\textwidth]{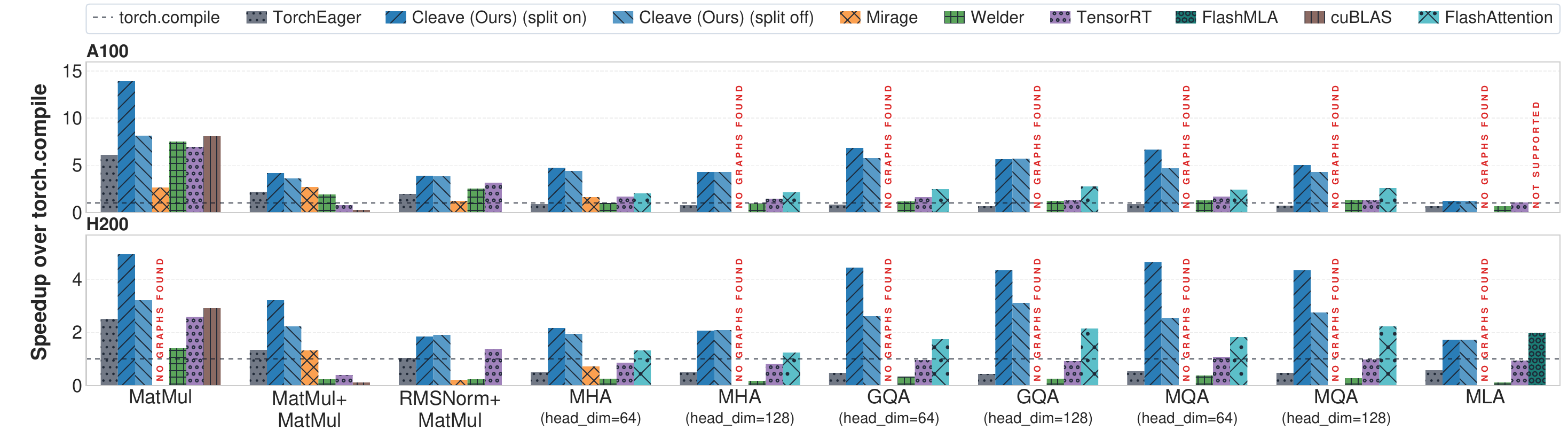}
  \caption{
  Decode-based subgraph-level kernel performance speedup against \texttt{torch.compile} with Torch eager mode, Mirage, Welder, TensorRT, cuBLAS, FlashAttention (which automatically uses FlashDecoding for small shapes), FlashMLA and \sysname (ours).}
  \label{fig:decode_operator_performance}
\end{figure*}

We evaluate \sysname and answer the following questions:

\begin{enumerate}[leftmargin=*]

\item \textbf{End-to-end impact:} Do \sysname-optimized kernels improve the performance of the transformer layer for a representative set of LLMs?

\item \textbf{Subgraph performance:} On fixed-shape subgraphs, can \sysname discover kernel optimizations that require algebraic transformations and operator scheduling? We answer this question for a variety of common subgraphs.

\item \textbf{Dynamic operator performance:} When a subgraph's shapes are known only at runtime, can \sysname compile it once and reach the performance of handwritten kernels?

\item \textbf{Compilation scalability:} How long does \sysname take to optimize kernels? How does compilation time change across subgraphs of different complexity?

\item \textbf{Ablation study:}  How does each part of our design affect the performance of \sysname-optimized kernels?
\end{enumerate}

We first describe the experimental setup, then present end-to-end and subgraph-level results, analyze compilation time and workload coverage, and finally study the effect of \sysname's main optimization components through ablations.

\subsection{Experimental Setup}\label{subsec:setup}

We evaluated \sysname using two single-GPU servers, one with an NVIDIA A100-SXM4-40GB and the other with an H200-SXM5-141 GB. Both servers have 12 CPU cores and 96GB of host memory. Our experiments were run on Ubuntu 22.04 and with CUDA 12.8.

\mypara{Workloads.} Our evaluation covers both subgraph-level and end-to-end workloads. At the subgraph level, we benchmark 9 subgraphs, including those that are commonly used in transformer models and those that have consecutive reduction nodes. For our end-to-end evaluations, we use transformer blocks from GPT-2 \cite{gpt2}, Llama-3 8B \cite{llama3}, and Qwen3-32B \cite{qwen3}. Across all benchmarks, computations use FP16 inputs, FP32 accumulation, and FP16 outputs. For the dynamic operators we use nine paged and ragged attention operators whose shapes are taken from workloads captured in \texttt{flashinfer-trace} from FlashInfer-Bench \cite{flashinferbench}, 303 in total, on the NVIDIA’s SOL-ExecBench~\cite{sol-execbench}.

\mypara{Baselines.} We compare our system against Mirage~\cite{mirage}, Welder~\cite{welder}, TensorRT~\cite{tensorrt}, \texttt{torch.compile}~\cite{ansel2024pytorch}, FlashAttention~\cite{flashattention}, FlashMLA~\cite{flashmla}, and cuBLAS~\cite{cublas} (see Appendix~\ref{s:alg-transform} for precise version).
For the dynamic operators from FlashInfer-Bench we compare against FlashInfer on its cuDNN~\cite{cudnn} and handwritten FA2 and FA3 backends~\cite{flashinfer}.

\mypara{Metrics.} To measure both search and subgraph/model performance, we report compilation and run time. Compilation time results are included for the search-based systems, namely \sysname, Mirage, and Welder, measuring the total CPU time including superoptimization, scheduling, and \linebreak NVCC compilation, while runtime measures the generated kernel latency using CUDA events on both individual subgraphs and end-to-end model execution. The dynamic operators are timed instead under NVIDIA's SOL-ExecBench harness~\cite{sol-execbench}, which FlashInfer-Bench~\cite{flashinferbench} is part of.

\subsection{End-to-end Benchmark}

\begin{figure}[h]
  \centering
  \includegraphics[width=0.45\textwidth]{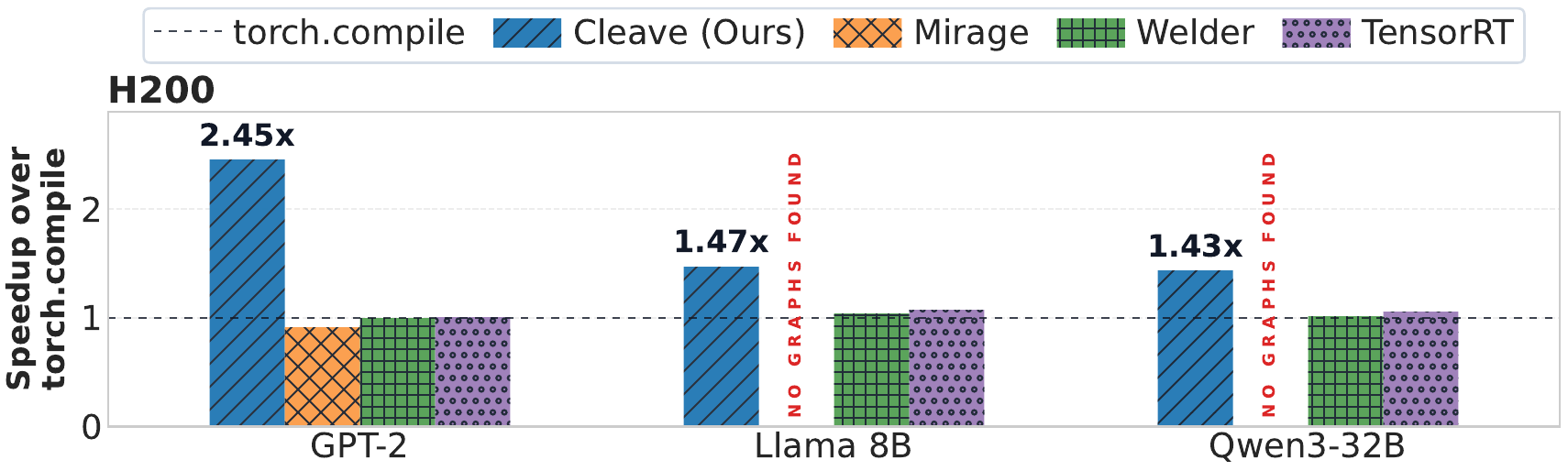}
  \caption{End-to-end performance on a single transformer layer across 3 different model architectures on H200 GPU.
  }
  \label{fig:end_to_end}
\end{figure}

For each evaluated architecture, we compare kernels generated by \texttt{torch.compile}, Mirage, Welder, TensorRT, and \sysname (ours) on a single transformer layer. Figure \ref{fig:end_to_end} reports the end-to-end execution speedup over \texttt{torch.compile} across multiple prefill shapes.

\sysname is the fastest system on all three models, achieving 2.45$\times$, 1.47$\times$, and 1.43$\times$ speedup over \texttt{torch.compile} on GPT-2, Llama-3 8B, and Qwen3-32B, respectively. TensorRT provides smaller gains on the larger models, while Welder is close to parity with \texttt{torch.compile}. Mirage cannot fully support the GQA-based executions used by Llama-3 8B and Qwen3-32B, so we only report Mirage results where execution is available.

Llama-8B and Qwen3-32B see smaller gains than GPT-2 because much of the computation time in these larger models is spent computing linear layers (\eg MLP Matmul). The baseline uses highly tuned vendor kernels for these operations, limiting the possible gains from compiler optimizations.

\subsection{Subgraph-level Benchmark}\label{subsec:operator-val}

We evaluated improvements on 9 subgraphs (Table~\ref{tab:compilation_time}). These include
(a) four non-causal attention variants: Multi-Head Attention (MHA), Group-Query Attention (GQA), Multi-Query Attention (MQA), and Multi-Head Latent Attention (MLA);
(b) three commonly used subgraphs that represent the common dense computation in transformer blocks: MatMul, RMSNorm+MatMul, and SwiGLU. In particular, optimizations from \flashattention can be applied to RMSNorm+MatMul, and horizontal fusion can be used with SwiGLU because it has a parallel reduction structure with a common input. We use shapes from Llama-3 8B for these subgraphs; (c) two consecutive reduction-based dense subgraphs where iterative tiling and memory planning are important for optimization: MatMul+MatMul and MatMul+RowSum; and (d) a subgraph containing a single MatMul operator, representing the base case where no optimization is performed.

We evaluate these subgraphs in two settings: prefill and decode. The prefill-setting uses large sequence- and context-lengths, while the decode-setting uses small sequence-lengths but has a large context-length. \splitk is particularly important for the decode-setting.\footnote{We report the shapes used for each subgraph in Appendix~\ref{s:alg-transform}.}

\begin{figure*}[t]
  \centering
  \includegraphics[width=1.0\textwidth]{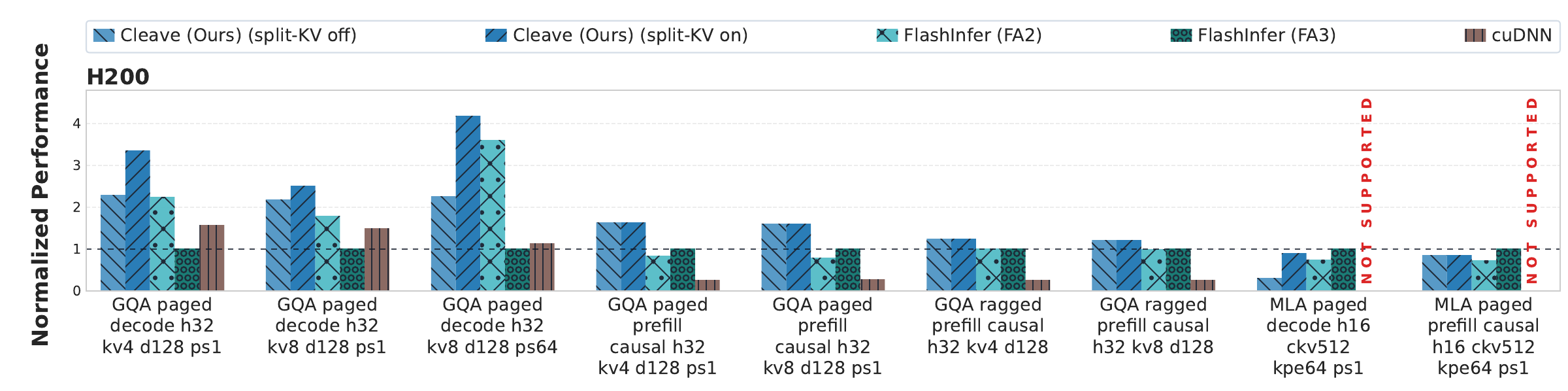}
    \caption{Dynamic-shape operator performance when toggling split-KV, against FlashInfer's FA2 and FA3 backends and cuDNN, normalized per operator to FA3 (dashed line at 1.0). Each bar is a geometric mean over the operator's captured workloads.
    Shapes have \emph{h} query heads, \emph{kv} key/value heads, \emph{d} head dimension and \emph{ps} page size; MLA gives \emph{ckv} and \emph{kpe}, the widths of the compressed KV cache and the RoPE cache.
    }
  \label{fig:eval-dynamic}
\end{figure*}

\mypara{Prefill setting.} Figure \ref{fig:prefill_operator_performance} compares the execution performance of \sysname optimized subgraphs to those optimized by the baseline.
 We evaluate all subgraphs using prefill-oriented shapes where both the input sequence length and context length are large ($\geq$ 256).
We report geometric mean speedup (across multiple shapes) compared to \texttt{torch.compile}.

We find that \sysname-optimized kernels have up to \maxOpRunSpeedup better performance than the best baseline on each individual subgraph. This is because \sysname's approach of combining algebraic transformation and operator scheduling allows it to optimize kernels across different shapes and different subgraphs. Consequently, in many cases \sysname kernels outperform even those built using the vendor-provided cuDNN library.
Finally, we found that Mirage could not find optimized graphs for subgraphs with large input shapes. For instance, it failed to find an optimized graph for MHA with head-dimension of 128, but could find one with head dimension of 64. This is because Mirage's superoptimization algorithm is shape-sensitive. By contrast, \sysname's symbolic approach allows it to scale to even these larger sizes.

\mypara{Decode setting.} Similarly, Figure~\ref{fig:decode_operator_performance} evaluates performance for decode-oriented shapes where the input sequence length or batch size remains small (< 256) while the context length is large ($\geq$ 1024).
\splitk is particularly important in this setting, and so we also present results when \sysname does not use \splitk (`split off'). Similar to the prefill setting, the results in this case demonstrate \sysname's effectiveness.
Beyond that, the results isolate the benefit of \splitk: with \splitk, \sysname generates kernels up to $\splikimprovement{} \times$ faster than without it.

For all three evaluated combinations, \Welder does not fuse consecutive reduction-based subgraphs once sequence length reaches 512. Consequently, in contrast to \sysname-optimized kernels, the \Welder-optimized kernels need to materialize intermediate tensors in GMEM, leading to significantly worse performance for long sequences.

\subsection{Dynamic Operator Benchmark}

We evaluate nine operators (see \S\ref{subsec:setup}) whose shapes are known only at runtime, covering paged and ragged attention, on the workloads captured for each of them in FlashInfer-Bench~\cite{flashinferbench}. Every operator is compiled once, and that single kernel runs all of its captured workloads, 303 in total and between 15 and 48 per operator, with no recompilation. Figure~\ref{fig:eval-dynamic} reports the result. \sysname is faster than FlashInfer's FA2 backend on all nine operators, by $1.16\times$ to $2.04\times$ (\maxOpRunSpeedupDynamicFAII geometric mean), and faster than its FA3 backend on seven of the nine (\maxOpRunSpeedupDynamicFAIII geometric mean); it is slower only on the two MLA operators ($0.85\times$--$0.89\times$), where FA3 merges its partial results inside a single cooperative kernel while \sysname pays a second launch. On the four decode operators, applying Split-K along the key-value axis of the scheduled graph is worth a further $1.15\times$ to $2.99\times$. However it is not beneficial for the remaining five (prefill) operators, whose grid is already filled by one work tile per query tile per request.

\subsection{Search Space}

Table \ref{tab:compilation_time} shows that \sysname achieves substantially better compilation scalability than prior automated kernel generation systems while maintaining broad workload coverage. With \splitk disabled, \sysname's compilation time (1--5 minutes) is comparable to \Welder's (1--4 minutes), while \Welder still misses important optimizations due to its lack of algebraic reasoning. With \splitk enabled, \sysname compiles each subgraph in 8--14 minutes on average. In contrast, \Mirage takes 121--224 minutes on average on the four subgraphs for which it generates valid code, and runs out of memory on two others. Furthermore, the version of \Mirage we used is unable to generate code for GQA, MQA, and MLA.

\begin{table}[h]
    \centering
    \small
    \begin{threeparttable}
        \begin{tabular}{@{}lcccccc@{}}
            \toprule
            \textbf{Subgraph} & \textbf{$n$} & \textbf{\Welder} & \multicolumn{2}{c}{\textbf{\Mirage}} &
            \multicolumn{2}{c}{\textbf{\sysname}} \\
            \splitk & & N/A & Off & On & Off & On \\
            \midrule
            MatMul          & 1 & 1 & 3 & 180   & 1 & 13 \\
            Matmul+MatMul   & 2 & 1 & 10 & 224  & 2 & 14 \\
            Matmul+Rowsum   & 2 & 2 & 4 & 209 & 5 & 12 \\
            SwiGLU          & 4 & 4 & 11 & OOM & 3 & 12 \\
            RMSNorm+MatMul  & 7 & 2 & 7 & OOM   & 2 & 9 \\
            MHA         & 7 & 2 & \xmark  & 121  & 3 & 8 \\
            GQA      & 7 & 3 & \xmark & \xmark & 3 & 8 \\
            MQA        & 7 & 2 & \xmark & \xmark & 3 & 8 \\
            MLA  & 7 & 3 & \xmark & \xmark & 3 & 9 \\
            \bottomrule
        \end{tabular}
        \caption{Compilation time (in minutes) of different subgraphs averaged across different shapes on A100. For \Mirage, we only average compilation times for cases where it generated valid code.
        \xmark\ means no valid code is generated for the given subgraph. $n$ denotes the number of nodes for the given subgraph.%
        }
        \label{tab:compilation_time}
    \end{threeparttable}
\end{table}

\subsection{Ablation Study}

Finally, we perform ablation studies to understand the contribution of each technique to the kernel performance.%

\begin{figure}[h]
  \centering
  \includegraphics[width=0.45\textwidth]{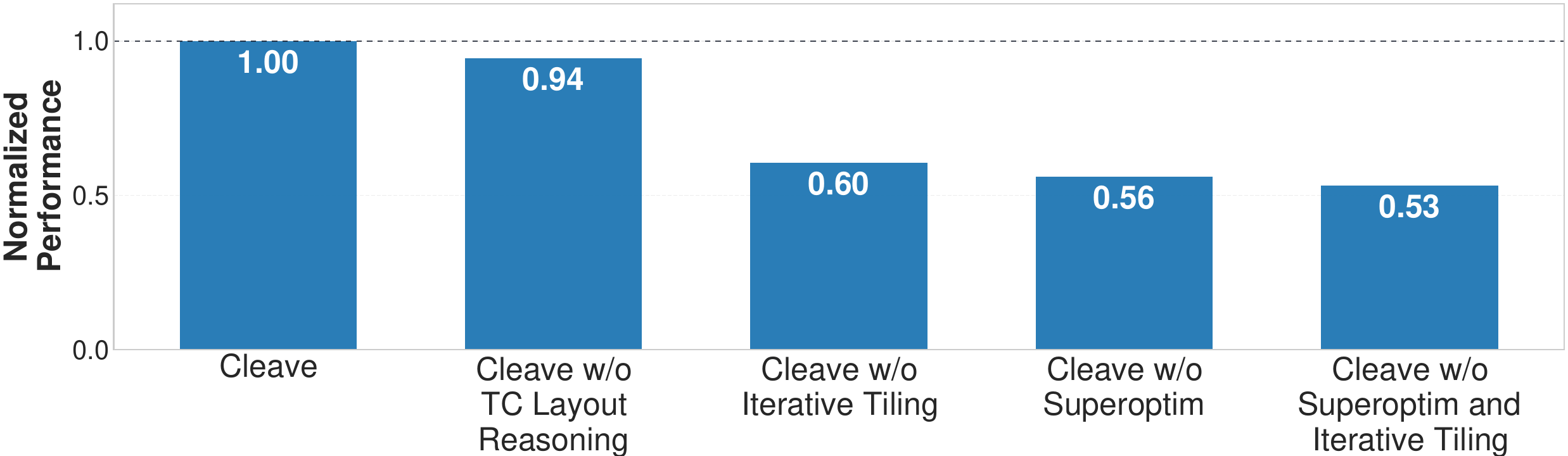}
  \caption{Component ablation study. Each \sysname bar removes \textbf{only} the component(s) listed. Y-axis shows the geometric mean of the normalized performance against the original \sysname implementation, across 9 subgraphs on various shapes.}
  \label{fig:component_ablation}
\end{figure}

In Figure~\ref{fig:component_ablation} we show the impact on kernel performance when different \sysname components are disabled. We report the geometric mean of normalized performance (normalized to that of kernels optimized using an unmodified \sysname) measured across 9 subgraphs and a variety of shapes. We observe that not (explicitly) maintaining tensor core layout constraints (\S\ref{s:codegen:memory} and Appendix~\ref{apx:tc})
leads to 6\% lower performance; changing scheduling so that iterative tiling schedules are not generated reduces performance by 40\%; and disabling algebraic transformation leads to a 44\% degradation. Finally, disabling both operator scheduling and algebraic transformation leads to a 47\% degradation. These results show the importance of combining algebraic transformation and scheduling: both contribute substantially to the final performance. In Appendix~\ref{s:alg-transform} we also study the effect of adding algebraic transformations to existing compilers.

\section{Related Work}

Prior works that automatically fuse and optimize kernels fall into the following broad categories:

\mypara{Graph and algebraic transformation.}
These compilers use graph substitution to search for transformations \cite{egg-mlsys, einnet, pet, korch, flashtensor, taso} to achieve several goals, including optimizing computation, minimizing intermediate tensor size, and increasing the number of operators that can be mapped to efficient vendor-provided implementations. The rules used by these compilers generally do not consider scheduling,
and miss many of the optimizations found by \sysname.

\taso uses superoptimization to automate the generation of substitution rules. However, to do this effectively, it must search through the set of all rules that can translate between any pair of equivalent graphs. This search space is large, and thus the rules it produces are constrained to small graphs. By contrast, we focus on finding graphs equivalent to a single input graph,
allowing us to scale to larger graphs.

\mypara{Scheduling-based tensor compilers.}
Several scheduling-based compilers have been proposed that aim to automate kernel fusion~\cite{welder, ladder, astitch, ansor, chimera, mcfuser, neptune, plus, roller, rammer}. These use a variety of approaches: AStitch~\cite{astitch}, Welder~\cite{welder}, and Ladder~\cite{ladder} introduce SMEM caching to automatic kernel fusion. However, they lack support for iterative tiling and thus the schedules they produce (which resemble inter-threadblock schedules) have large SMEM usage. On the other hand, Ansor~\cite{ansor} uses evolutionary search to find a schedule built using TVM's scheduling primitives. This approach is general, but slow (and unscalable) due to the large search space. By contrast,  \sysname uses a cost-model-guided search that achieves fast search time while discovering high-performance kernels.  Chimera~\cite{chimera} and MCFuser~\cite{mcfuser} adopt a simpler approach to perform iterative tiling by identifying loops with identical access patterns. However, this simple analysis cannot optimize more general kernels, e.g., ones where inner loops are different.
Recent scheduling-based compilers, including \Neptune~\cite{neptune}, Nautilus~\cite{zhao2026nautilus} and Flashlight~\cite{you2026flashlight} add a scheduling pass for online updates (as used by \flashattention), that can fuse two loops with attention-like reduction and dependency structure.
In doing so, they use expert-provided algebraic transformation rules, and thus, unlike \sysname, cannot automatically discover algebraic transformations. Extending this approach to
other transformations, including \splitk is challenging. %
Importantly, none of the above solves the challenge of combining scheduling with general algebraic transformations and scaling search.

\mypara{Joint optimization.} Full-stack tensor compilers today, such as TVM~\cite{tvm}, ONNXRuntime~\cite{onnxruntime}, and TorchInductor~\cite{ansel2024pytorch},
also conduct both algebraic transformations and scheduling. However, they mostly rely on rule-based graph rewrite for algebraic optimization and expert-written scheduling templates, limiting their ability to find optimizations.

As we discussed previously (\S\ref{sec:introduction}), \Mirage~\cite{mirage} uses superoptimization to optimize algebraic structure and scheduling. As \S\ref{sec:evaluation} shows, this leads to too large a search space and limits its ability to optimize common kernels. %
Prism~\cite{wu2026prism} extends \Mirage by making the tile sizes of candidate kernels (\ie grid dimensions and loop ranges) symbolic to optimize search efficiency. %
However, it continues to explore algebraic structure and the scheduling decisions of which dimension to tile jointly, which can still lead to a large search space. %

Concurrent with our work, \Trinity~\cite{trinity}, proposed a stateful loop-based IR and adapted e-graph algorithms to perform algebraic transformations, scheduling, and memory optimizations jointly. This approach can discover fine-grained joint optimizations, e.g., online softmax, which \sysname, Mirage and other existing compilers cannot (except for Neptune~\cite{neptune}). However, like other substitution-based approaches, it requires handwritten rules and does not currently support searching for loop tiling. Instead, users need to provide a tiled program as input, and therefore require additional user effort to recover \sysname's optimizations.

\mypara{Agentic kernel optimization.} Using LLMs to optimize kernels~\cite{kda,autogpukernel,accelopt,cake,ksearch,kernelevolve} has become an active area of research: one or multiple LLM-based agents propose and revise kernel code and select candidates by compiling, testing, and timing them. Although they can match or outperform expert-written kernels on some workloads, %
it often consumes a substantial amount of tokens: CAKE, for example, reports a budget of 80 million tokens for a single optimization run~\cite{cake}. This cost recurs for every new optimization problem, which makes these approaches still expensive to use widely.

\section{Conclusion}
\sysname demonstrates that decoupling algebraic transformation and operator scheduling is key to enabling better ML kernel optimizations. It allows compilers to optimize and produce high-performance kernels within reasonable time.
Its algebraic transformation works on graphs with symbolic tensor shapes and reducibility-pruned search space in order to scale enumeration-based search. Its operator scheduling performs iterative tiling and horizontal fusion to fuse across reduction operators and reduce SMEM usage. Our evaluation shows that \sysname reduces compilation time by $5.9\times$ on average compared to the joint-search compiler, \Mirage, while generating kernels up to $2.8\times$ faster than the best baseline.

\begin{acks}
This work was supported in part by grants from Google, and AMD; and was done using resources, services and staff expertise provided by NYU IT High Performance Computing. 
\end{acks}

\bibliographystyle{ACM-Reference-Format}
\bibliography{main}
\newpage
\appendix
\section{Implementation Details}
\label{sec:implementation}

\mypara{Symbolic graph conversion.}
\sysname uses a simple implementation for graph symbolization (\S\ref{sec:algebraic-transformation}). It runs in two passes: a \emph{grouping pass} that analyzes the operator constraints and merges dimensions constrained to have the same size, and a \emph{symbolization pass} that allocates a distinct symbol to dimensions in each group.
The grouping pass traverses each operator in the graph and applies the corresponding grouping rule as listed in Table~\ref{tab:symbolic-grouping}. Reduction dimensions are grouped together, and batch dimensions are grouped when they match without broadcasting. Any dimension of size $1$ is placed in a dedicated broadcast group so that the \algebraic graph preserves the original broadcasting pattern. In the symbolization pass, all dimensions in the broadcast group are assigned the value $1$, and each remaining group receives a distinct symbol.
For example, Matmul$([8, 32,64] \times [1,64,128] \to [8,32,128])$ will be distilled into Matmul$([b,m,k] \times [1,k,n] \to [b,m,n])$.\footnote{We use batched Matmul where the batch dimension can broadcast.}

\begin{table}[H]
\centering
\footnotesize
\setlength{\tabcolsep}{5pt}
\renewcommand{\arraystretch}{1.1}
\begin{tabularx}{\columnwidth}{@{}l >{\raggedright\arraybackslash}X@{}}
\toprule
\textbf{Operator} & \textbf{Grouping rules} \\
\midrule
$C=\operatorname{MatMul}(A,B)$
& $A_k \sim B_k$, \ $A_m \sim C_m$, \ $B_n \sim C_n$; batch axes are grouped pairwise when no broadcasting occurs. \\

$C=\operatorname{ElemBinary}(A,B)$
& For each axis $j$, $A_j \sim B_j \sim C_j$ when no broadcasting occurs. \\

$O=\operatorname{ElemUnary}(I)$
& For each axis $j$, $I_j \sim O_j$ when no broadcasting occurs. \\

$O=\operatorname{Reduction}_r(I)$
& For each axis $j \neq r$, $I_j \sim O_j$; axis $r$ forms a separate group. \\
\bottomrule
\end{tabularx}
\caption{
Dimension grouping rules.
$A_j$ denotes the $j$-th dimension of tensor $A$. $A_j \sim B_j$ means the two dimensions are grouped together and receive the same symbol.
Axes of extent $1$ form a broadcast group and are not merged with other axes.
}
\label{tab:symbolic-grouping}
\end{table}

\mypara{Supporting \algebraic graph superoptimization.}
We extend \Mirage to support superoptimization on an \algebraic graph. The search algorithm is adapted in a few places. First, when inserting a new operator, the shape validity checking on the input and shape inference of the output should be extended to operate on the symbolic shape space. As the symbols can be instantiated with any arbitrary value, we consider the insertion valid only when it is valid for all instantiations. For efficiency, we check validity by assigning each symbol a concrete and distinct value and checking on the concrete domain. Though this may fail to reject some invalid graphs, it does not affect soundness, as it is guarded by the probabilistic checker.
Second, the probabilistic testing requires running on concrete dimensions, for which we instantiate a graph with a distinct and \emph{small} value per symbol and perform the test on the small graph. This is the core to efficient testing.
Third, we restrict the search to one-level (\ie no block graphs from \Mirage), as our superoptimization involves kernel scheduling only in \splitk, which is resolved with our \Split operator.

\mypara{Tile Size Proposal}
\sysname adopts a framework to systematically optimize data flow execution on hardware accelerators. Specifically, the algorithm first isolates hardware-imposed reduction constraints, then explores a search space of valid iterative graph configurations by grouping compatible reductions. Algorithm \ref{alg:cleave_itile_generation} details this procedure.

\begin{algorithm}[t]
\caption{Tile size configuration generation.}
\label{alg:cleave_itile_generation}

\KwIn{Concrete transformed graph $G_t$}
\KwOut{Tile size configuration set $\mathrm{TileConfigs}$}

\SetKwFunction{Cleave}{EnumerateTileSizes}
\SetKwFunction{MinRed}{GetMinReductionRange}
\SetKwFunction{EnumOut}{EnumerateOutputTiles}
\SetKwFunction{Group}{GroupedReductionConfigs}
\SetKwFunction{Schedule}{itGraphScheduling}

\Fn{\Cleave{$G_t$}}{

    \tcp{Step 1: infer reduction constraints from hardware specifications}
    $R \leftarrow \emptyset$\;

    \For{$n \in G_t$}{
        \If{$n$ is a reduction-based op}{
            $R[n] \leftarrow \MinRed(n)$\;
        }
    }

    \tcp{Step 2: generate iterative tiling candidates}
    $\mathrm{TileConfigs} \leftarrow \emptyset$\;

    \For{$T \in \EnumOut(G_t, R)$}{
        \For{$c \in \Group(G_t, T, R)$}{
            $\mathrm{TileConfigs} \leftarrow \mathrm{TileConfigs} \cup \{c\}$\;
        }
    }

    \Return{$\mathrm{TileConfigs}$}\;
}
\end{algorithm}

\begin{itemize}[leftmargin=*]
\item \textbf{Step 1 (lines 2---5):} Analyze hardware constraints and build base iterations. First, given a concrete transformed graph $G_t$, the scheduler initiates a constraint analysis to identify limitations on reduction operations. \sysname iterates through the graph nodes: for every node identified as a reduction, the algorithm invokes the \linebreak \texttt{GetMinReductionRange} function to populate a reduction constraint map, $R$. This step is crucial for determining the mandatory lower-bound ranges imposed by specific hardware units, such as Tensor Core or SIMT coalescing-aware dimensions. %

\item \textbf{Step 2 (lines 6---10):} Then, the scheduler generates candidate iterative graphs. \sysname utilizes Roller's approach to produce the set of potential output tile shapes in the \texttt{EnumerateOutputTiles} interface. This is achieved by expanding a base output tile into several larger valid output tile shapes. Such shapes will act as candidate output tiles in our system.
However, each output tile candidate is further enhanced with reduction metadata. Subsequently, for each tile shape $T$, \sysname calls \texttt{GroupedReductionConfigs} to explore various reduction tile sizes, \ie the tiling dimensions on the reduction axes. To later facilitate the scheduling, this exploration groups reduction nodes to enumerate together, using the same rule as intra-threadblock scheduling, as they will share the same reduction tile size.
\end{itemize}

\mypara{\itgraph cost model.}
\sysname employs an analytical cost model that analyzes the cumulative GMEM traffic across all iterations of the \itgraph, and the peak SMEM usage across the execution. Currently, our cost model assumes all buffers are located in SMEM for simplicity. Specifically, \sysname simulates the execution of the \itgraph on a single threadblock and tracks the memory allocation with a fake SMEM allocator. This allows \sysname to account for the buffer liveness and the reuse of SMEM across different operators, which is crucial for accurately estimating the peak SMEM usage. For GMEM traffic estimation, \sysname accumulates the buffer sizes of Input and Output nodes that access GMEM across all inner-loop iterations.

\section{Maintaining Tensor Core Constraints}\label{apx:tc}
\begin{figure}[h]
  \centering
  \includegraphics[width=1.0\linewidth]{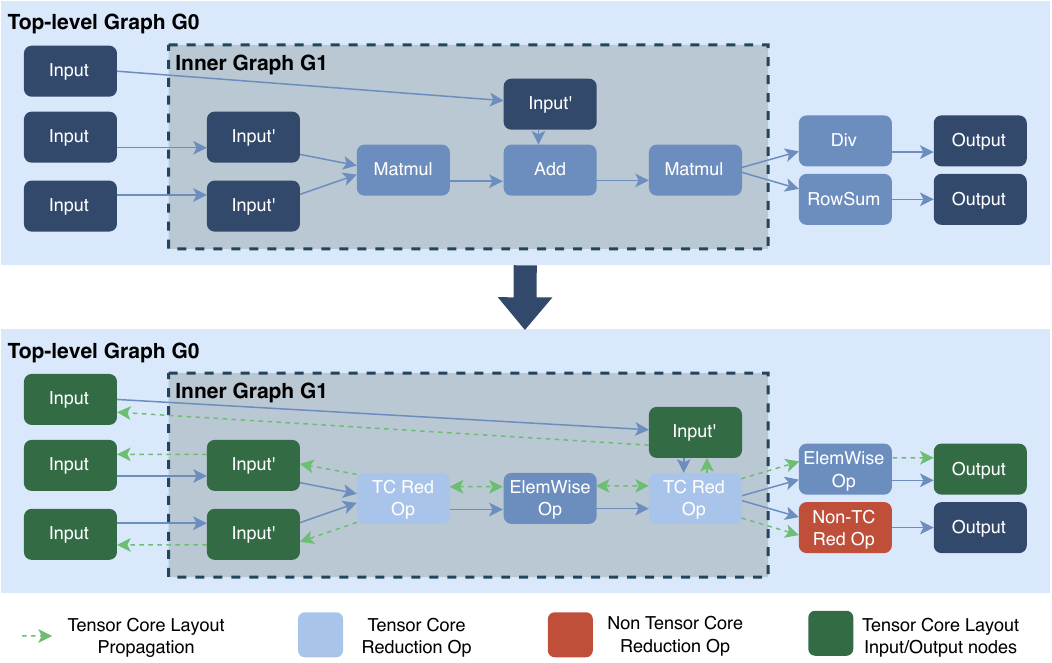}
  \caption{Multi-level Tensor Core Layout awareness propagation through the \itgraph.}
  \label{fig:cleave_tc_layout_prop}
\end{figure}

Tensor cores impose layout constraints. To maintain these, \sysname propagates the constraints through the \itgraph starting from MatMul nodes, extending across element-wise operators and stopping at non-element-wise boundaries. This identifies buffers that need to be represented with tensor core fragment layouts. Figure \ref{fig:cleave_tc_layout_prop} illustrates such tensor core layout propagation over a hierarchical \itgraph with two tensor core matrix multiplications inside the nested inner graph. For such buffers, when writing to GMEM, \sysname reorganizes data through SMEM to enable vectorized and coalesced transfers; for input buffers loaded from GMEM, it applies swizzled SMEM layouts to avoid bank conflicts.

\section{Additional Evaluation}
\label{s:alg-transform}

\mypara{Codegen.} Throughout our evaluation we use our TileLang codegen backend with version v0.1.5.

\mypara{Baseline version details.} We compare our system \sysname against Mirage with commit hash \texttt{e8980e1} \cite{mirage}, Welder with commit hash \texttt{af53ab1} \cite{welder}, \texttt{torch.compile} (Inductor) v2.9.0+cu128 \cite{ansel2024pytorch}, TensorRT v10.15.1.29 \cite{tensorrt}, the original FlashAttention v2.8.3 \cite{flashattention}, FlashMLA with commit hash \texttt{47c35a7} \cite{flashmla}, and cuBLAS v12.8 \cite{cublas}.%

\newcolumntype{L}[1]{>{\raggedright\arraybackslash}p{#1}}

\begin{table}[H]
    \centering
    \small
    \setlength{\tabcolsep}{4pt}
    \begin{threeparttable}
        \begin{tabular}{@{}lL{0.68\columnwidth}@{}}
            \toprule
            \textbf{Subgraph} & \textbf{Prefill Shapes} \\
            \midrule
            MatMul & $X:(s,4096),\ W:(4096,4096)$ \\
            \midrule
            MatMul+MatMul & $Q:(32,s,64),\ K,V:(32,L,64)$ \\
            \midrule
            MatMul+RowSum & $X:(s,64),\ W:(64,4096)$ \\
            \midrule
            SwiGLU & $X:(s,4096),\ W_1,W_2:(4096,14336)$ \\
            \midrule
            RMSNorm+MatMul & $X:(s,4096),\ W:(4096,6144)$ \\
            \midrule
            MHA & $Q:(32,s,d),\ K:(32,d,L),\ V:(32,L,d)$ \\
            \midrule
            GQA & $Q:(32,s,d),\ K:(8,d,L),\ V:(8,L,d)$ \\
            \midrule
            MQA & $Q:(32,s,d),\ K:(1,d,L),\ V:(1,L,d)$ \\
            \midrule
            MLA & $Q:(128,s,dr),\ K:(128,L,dr),\linebreak
            V:(128,L,d)$ \\
            \bottomrule
        \end{tabular}
        \caption{Shapes used in the prefill subgraph benchmark. We sweep $s \in \{256, 512, \ldots, 4096\}$. For all attention subgraphs, $L=4096$. For MHA, GQA, and MQA, $d \in \{64,128\}$. For MLA, $d=128$ and $dr=192$.}
        \label{tab:prefill_shapes}
    \end{threeparttable}
\end{table}

\mypara{Prefill and decode subgraph shapes.}  Tables~\ref{tab:prefill_shapes} and~\ref{tab:decode_shapes} summarize the tensor shapes used in our prefill and decode subgraph benchmarks (\S\ref{subsec:operator-val}).

We use the following notation: Across all subgraphs, $X$ denotes the input activation, $W$ denotes a weight matrix, and $Q$, $K$, and $V$ denote the query, key, and value tensors;
$s$ denotes the prefill sequence length, while $b$ denotes the decode batch size.
For attention subgraphs,  $L$ denotes the context (or KV-cache) length, and $d$ denotes the per-head hidden dimension, and the leading dimensions (e.g., $32$ or $8$) show the number of query or KV heads used by the attention variant.
For MLA,
$dc$ denotes the content dimension on the absorbed variation of MLA \cite{liu2024deepseek}, $dr$ denotes the positional/rotary dimension, and $Q_{pe}$ and $K_{pe}$ denote the positional-encoding components of the query and key.

\begin{figure*}[t]
  \centering
  \includegraphics[width=0.85\textwidth]{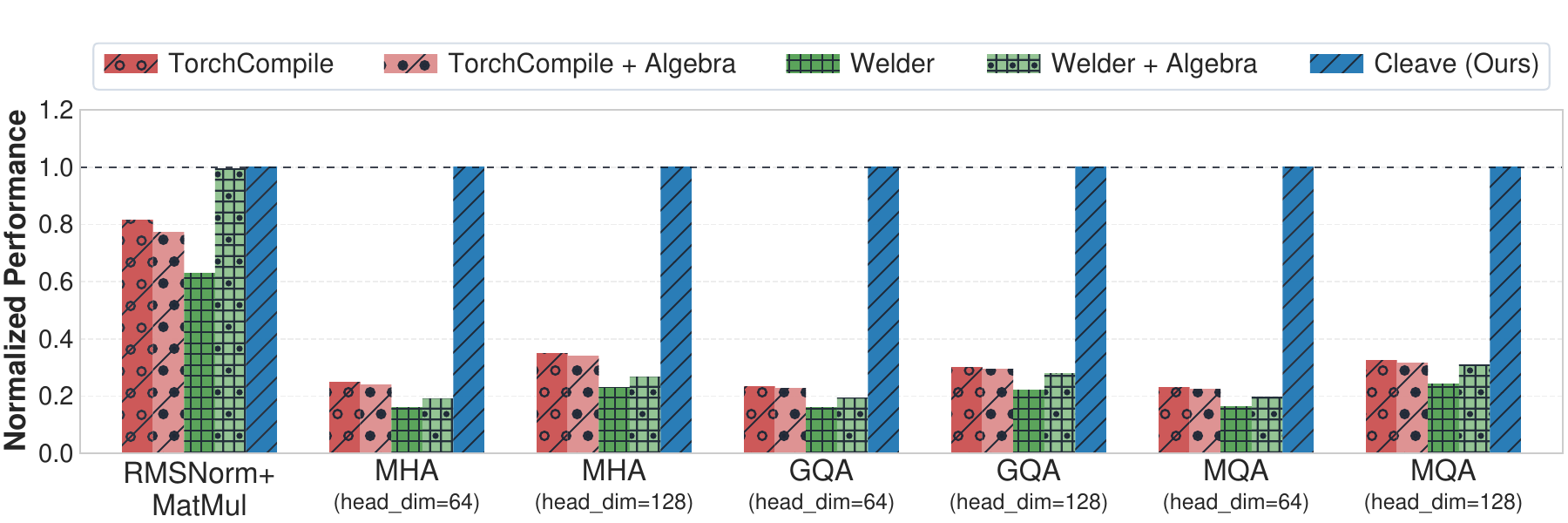}
  \caption{Providing algebraic transformations to \texttt{torch.compile} and Welder.}
  \label{fig:algebra_ablation}
\end{figure*}

\begin{table}[H]
    \centering
    \small
    \setlength{\tabcolsep}{4pt}
    \begin{threeparttable}
        \begin{tabular}{@{}lL{0.68\columnwidth}@{}}
            \toprule
            \textbf{Subgraph} & \textbf{Decode Shapes} \\
            \midrule
            MatMul & $X:(b,4096),\ W:(4096,b)$ \\
            \midrule
            Matmul+MatMul & $Q:(32,b,64),\ K,V:(32,L,64)$ \\
            \midrule
            Matmul+Rowsum & $X:(b,64),\ W:(64,4096)$ \\
            \midrule
            SwiGLU & $X:(b,4096),\ W_1,W_2:(4096,14336)$ \\
            \midrule
            RMSNorm+MatMul & $X:(b,4096),\ W:(4096,6144)$ \\
            \midrule
            MHA & $Q:(32,b,d),\ K:(32,d,L),\ V:(32,L,d)$ \\
            \midrule
            GQA & $Q:(32,b,d),\ K:(8,d,L),\ V:(8,L,d)$ \\
            \midrule
            MQA & $Q:(32,b,d),\ K:(1,d,L),\ V:(1,L,d)$ \\
            \midrule
            MLA & $Q:(128,128,dc),\ KV:(128,L,dc),\linebreak Q_{pe}:(128,128,dr),\ K_{pe}:(128,L,dr)$ \\
            \bottomrule
        \end{tabular}
        \caption{Shapes used in the decode subgraph benchmark. We sweep $b \in \{16,32,64,128\}$. For MHA, GQA and MQA, $d \in \{64,128\}$ and $L=4096$. For MLA, we sweep $L \in \{1024,2048,4096\}$, with $dc=512$ and $dr=64$.}
        \label{tab:decode_shapes}
    \end{threeparttable}
\end{table}

\mypara{Enabling algebraic transformations.} Next we evaluate whether our approach to algebraic transformation can help existing compilers. We do so by using \texttt{torch.compile} and Welder to compile graphs output by \sysname's algebraic transformation step (we used the best graph in all cases). Our results, in Figure~\ref{fig:algebra_ablation}, show the normalized performance (compared to \sysname) for 4 subgraphs and different shapes. We observe that \Welder has better performance when provided transformed graphs, though the kernels still perform worse than \sysname-optimized kernels. This demonstrates the benefit of algebraic transformation for scheduling compilers.

Surprisingly, algebraic transformation generally led to worse performance for \texttt{torch.compile}. We analyzed the TorchInductor log to understand the reason, and found that in this case changes made during algebraic transformation meant that the compiler was not able to identify fusion patterns (that it found in the original graph) and thus did not fuse all kernels. This demonstrates the challenge of using algebraic transformation with a rule-based compiler.

\begin{figure}[H]
  \centering
  \includegraphics[width=0.49\textwidth]{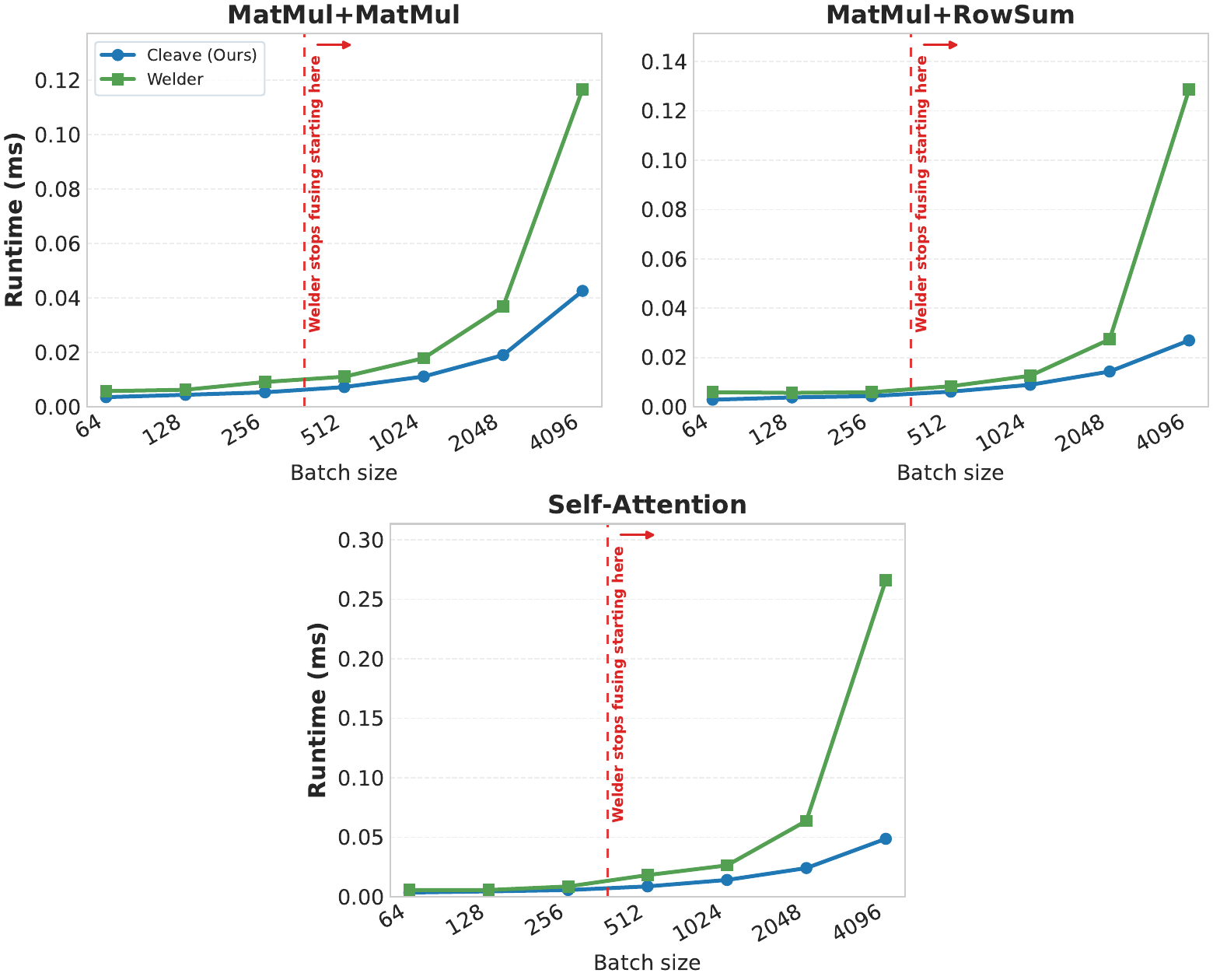}
  \caption{Performance comparison on three different subgraphs against Welder with increasing input sequence length.}
  \label{fig:welder_no_fusion_single}
\end{figure}

\mypara{Scaling fusion to long sequences.}
We also evaluated the effect of increasing input-sequence length on \sysname's scheduling pass. For this evaluation, we compared the performance of \sysname and \Welder optimized kernels for self-attention, MatMul + MatMul, and MatMul + RowSum. Figure~\ref{fig:welder_no_fusion_single} reports the runtime as sequence length increases on different workloads. We demonstrate one limitation of \Welder (discussed in \S\ref{s:bg:case}): it cannot fuse the attention graphs (or similar workloads that involve consecutive reductions, e.g., MatMul + MatMul and MatMul + RowSum) when the sequence length is 512 or larger, and thus exhibits significant performance degradation for longer sequence lengths.

\end{document}